\documentclass[journal=jpcafh,manuscript=article]{achemso}
\setkeys{acs}{usetitle = true}
\usepackage{subfig}
\usepackage{upgreek}
\usepackage{color}

\newcommand{\rad}{\hspace{0.3ex}\raisebox{0.8ex}{\tiny{\ensuremath{\bullet}}}}

\author{E. J. McBride}
\altaffiliation{Atomistic Simulation Centre}
\author{T. J. Millar}
\altaffiliation{Astrophysics Research Centre}
\author{J. J. Kohanoff}
\email{j.kohanoff@qub.ac.uk}
\altaffiliation{Atomistic Simulation Centre}
\affiliation[Queen's University Belfast]
{School of Mathematics and Physics, Queen's University Belfast, 
Belfast BT7 1NN, Northern Ireland, UK}

\title[\texttt{achemso} demonstration]
{The Irradiation of Water Ice by C$^+$ ions in the Cosmic Environment}

\begin{document}
\begin{abstract}
We present a first principles molecular dynamics (FPMD) study of the interaction of low energy, 
positively charged, carbon (C$^+$) projectiles with amorphous solid water clusters at 30 K. 
Reactions involving the carbon ion at an initial energy of 11 eV and 1.7 eV with 30-molecule 
clusters have been investigated. Simulations indicate that the neutral isoformyl radical, 
COH$\rad$, and carbon monoxide, CO, are the dominant products of these reactions. All these
reactions are accompanied by the transfer of a proton from the reacting water molecule to the
ice, where it forms a hydronium ion. We find that COH$\rad$ is formed either via a direct, ``knock-out'', 
mechanism following the impact of the C$^+$ projectile upon a water molecule or by creation of 
a COH$_2^+$ intermediate. The direct mechanism is more prominent at higher energies. CO is 
generally produced following the dissociation of COH$\rad$. More frequent production of the 
formyl radical, HCO$\rad$, is observed here than in gas phase calculations. A less commonly 
occurring product is the dihydroxymethyl, CH(OH)$_2\rad$, radical. Although a minor result, 
its existence gives an indication of the increasing chemical complexity which is possible in 
such heterogeneous environments. \\


\end{abstract}

\section{Introduction}
Water ice is an important, often dominant, component of molecular ices in cold interstellar 
clouds, in bodies in the Solar System and, by extension, in extra-solar planets and proto-planetary 
disks. Such ices can experience non-thermal irradiation and interactions by UV and X-ray photons, 
particularly in the neighborhood of young stars, by cosmic-ray particles and high-energy ions and 
electrons.  Particle acceleration can occur in shock fronts associated with supernovae explosions 
-- see Dumas \emph{et al}. \cite{Dumas2014} for a recent example of a supernovae remnant interacting 
with its surrounding molecular cloud --, in stellar flares, perhaps driven by episodic accretion in 
the formation of protostars, in Solar and stellar winds, and in planetary magnetospheres. It is thus 
of interest to investigate the interaction of water ice with abundant atoms and atomic ions.

Ions are crucial to the chemistry of the interstellar medium (ISM) \cite{Dalgarno1993, Herbst2001}. 
This is especially true in photon-dominated regions (PDRs) where interstellar gas is exposed to UV 
radiation from distant stars, leading to the formation of ionized species via photodissociation 
and photoionization. In these high-energy environments the carbon cation, C$^+$, is predicted 
to be the most abundant ion \cite{Lee1996}.

Carbon ions are also significant within our own solar system. Those found in planetary magnetospheres 
are believed to originate from the slow component of the solar wind \cite{Bochsler2000} while the 
Cassini Cosmic Dust analyzer has detected C$^+$ as a dominant component in the stream particles of 
Jupiter and Saturn \cite{Kempf2005}.

The reaction of C$^+$ with various small molecules has been the subject of much experimental and 
theoretical study \cite{Schiff1979, Herbst1983, Dalgarno1993, Herbst2001, MartinezJr.2008, Maergoiz2009}. 
While the majority of previous work has focused on understanding gas-phase chemistry, here we 
investigate this reaction in the condensed phase, within the ice which is present on planetary bodies 
and which also envelopes the microscopic dust particles that are ubiquitous in the ISM\cite{Herbst2009}. 

In fact, heterogeneous reactions are believed to be important in the synthesis of complex molecules. 
Purely gas-phase routes to complex molecules have been shown to be inefficient while also failing to 
reproduce the observed abundances of some simpler species. It is thought that the modification
and subsequent evaporation of ices in astrophysical environments by energetic particles may lead to 
the formation of larger organic species. One such example is the interaction of carbon ions with the 
icy surfaces present on Jupiter's Galilean moons. This reaction is evidenced by the detection of 
CO$_2$ on Callisto and Ganymede\cite{McCord1997}.

Essentially the same phenomenon occurs upon C$^{+q}$ irradiation of biological matter, when the ions 
gradually lose their energy to the medium along the radiation track, until their velocities are low 
enough that they can chemically react with water and other biomolecules. This happens just after 
the Bragg peak, at the end of the track, when ions have lost most of their kinetic energy and come 
close to rest.

Several studies have been undertaken to elucidate the mechanisms involved in reactions of C$^+$ with 
gas-phase water. Sonnenfroh \emph{et al.} \cite{Sonnenfroh1985} conducted one of the first experimental 
investigations of the $\textrm{C}^+ + \textrm{H}_2\textrm{O}$ reaction in the gas-phase, a crossed beam 
study with collision energies of 0.62 eV and 2.14 eV. They suggest that at the lower energy the 
production of either the formyl, HCO$^+$, or isoformyl ion, COH$^+$, occurs through the formation of 
a transient hydroxymethylene ion intermediate, HCOH$^+$. Subsequently, this complex decays via cleavage 
of the C-H bond to form the isoformyl cation COH$^+$ or by O-H bond cleavage to form the more stable 
formyl cation HCO$^+$,

\begin{equation}\label{eq:sonnenfroh}
 C^+ + H_{2}O \rightarrow HCOH^+ \rightarrow [HCO^+] + H\rad \qquad .
\end{equation}
In \ref{eq:sonnenfroh} the notation $[HCO^+]$ denotes both the formyl and isoformyl ions. 
Sonnenfroh \emph{et al.} reported that the formation of HCO$^+$ represents a significant fraction 
(20\% -- 30\%) of the final products at the lower energy.

At the higher collision energy the fraction of HCO$^+$ produced is found to decrease. Sonnenfroh 
\emph{et al.} proposed that this is the result of the increasing dominance of a direct, ``knock-out'' 
mechanism. In this alternative pathway the incoming C$^+$ cation strikes H$_2$O with an almost zero 
impact parameter causing the ejection of a hydrogen atom. The less stable isomer, COH$^+$, is 
formed exclusively here. 
Their observations also suggest that at the lower collision energy approximately one third of the 
products have sufficient energy to surmount the HCO$^+$/ COH$^+$ isomerization barrier. At the higher 
collision energy this is the case for two thirds of the products.

Ishikawa \emph{et al.} \cite{Ishikawa2001} performed an ab initio QCISD/6-31+G* direct molecular 
dynamics (MD) study of the reaction of C$^+$ with water and the mechanism of reaction was examined. 
This work represented an attempt to approximate the low energy beam studies of Sonnenfroh \emph{et al.}, 
the reactants here colliding with a relative kinetic energy of 0.62 eV. Their results have shown that 
the principal products are COH$^+$ and H$\rad$. 

\begin{equation}\label{eq:ishi1}
 C^+ + H_{2}O \rightarrow COH^+ + H\rad
\end{equation}

\begin{equation}\label{eq:ishi2}
 C^+ + H_{2}O \rightarrow COH_2^+ \rightarrow COH^+ + H\rad 
\end{equation}

They have also shown that the two main channels in this reaction are by direct ``knock-out'' of a 
hydrogen (see \ref{eq:ishi1}) and via an intermediate, COH$_2^+$ (see \ref{eq:ishi2}). 
This intermediate is not one which has previously been proposed in either experimental or theoretical 
studies. Most of the isoformyl cations produced were found to be internally energetic enough to 
isomerize to the formyl cation.

The C$^+$-water complex, COH$_2^+$, was found to be the most common intermediate (see 
\ref{fig:opt_coh2}). This is at odds with the work of Sonnenfroh \emph{et al.} who proposed 
that the hydroxymethylene intermediate, HCOH$^+$, should form. Ishikawa \emph{et al.} suggested 
that the hydroxymethylene cation had been considered to be significant based on observations that 
C$^+$ undergoes bond insertion reactions \cite{Ishikawa2001}. They stated, however, that these 
bond insertions had been with hydrocarbons which possess neither the polarity nor the convenient 
point of attachment presented by water. 

Ishikawa \emph{et al.} concluded that the collision of C$^+$ and H$_2$O at 0.62 eV produces COH$^+$, 
with HCO$^+$ as a rare product of the immediate reaction. The great majority of HCO$^+$ must come via 
isomerization and most COH$^+$ possesses sufficient internal energy to isomerize. Their findings were 
substantiated by further study \cite{Ishikawa2003}.

Flores \cite{Flores2006} has presented a study of the $\textrm{C}^+ + \textrm{H}_2\textrm{O} 
\rightarrow \textrm{COH}^+ + \textrm{H}$ reaction in which the quasiclassical trajectory method has 
been employed. The electronic structure has been calculated using a B3LYP-type density functional 
approach while the potential energy surface has been represented via a finite element method. 
Trajectory computations have been performed at a fixed relative translational energy of 0.62 eV that
corresponds to the crossed beam experiments of Sonnenfroh \emph{et al.} 

COH$_2^+$ is a critical intermediate here, in agreement with work by Ishikawa \emph{et al.} and in 
contrast with the interpretation of the crossed-beam experiments of Sonnenfroh \emph{et al.} Virtually 
all trajectories generate COH$^+$ but a significant proportion of the isoformyl cation is formed with 
enough vibrational energy to isomerize to HCO$^+$. Results here are almost fully coincident with 
Ishikawa. This work suggests that that the reason for the discrepancy in the relative translational 
energy distribution between the crossed beam experiments and the trajectory computations should be 
either tunnelling or the contribution from the excited states. 

Further analysis by Flores \emph{et al.} \cite{Flores2008} suggested that the first excited electronic 
state of COH$_2^+$ could be important in the generation of the formyl isomer which has been detected 
in crossed-beam experiments but not in quasiclassical trajectory computations. Later work indicated 
that tunnelling has little impact on COH$^+$/HCO$^+$ branching ratios \cite{Flores2009}. They showed 
that the lowest-lying excited state presents topological features which favor the production of 
HCO$^+$. Ab-initio molecular dynamics simulation of the $\textrm{C}^+ + \textrm{H}_2\textrm{O}$ 
reaction by Yamamoto \cite{Yamamoto2010} have reiterated many of the findings of Flores, Ishikawa 
and their co-workers.

Despite the perceived significance of this process, studies of carbon ion irradiation of water in 
the condensed phase have been quite limited. Strazzulla \emph{et al.}\cite{Strazzulla2003} have used 
30 keV $^{13}$C$^+$ ions to irradiate 1 $\upmu$m thick water ice films at 16 K and 77 K.
$^{13}$CO$_2$ was observed as the dominant product while a small quantity of $^{13}$CO was observed 
only at higher ion fluences. Dawes \emph{et al.}\cite{Dawes2007} have irradiated water ice at 30 and 
90 K using low energy 4 keV $^{13}$C$^+$ and $^{13}$C$^{2+}$ ions. $^{13}$CO$_2$ was observed as the 
only carbon bearing species formed. A later study \cite{Hunniford2009} using 2 keV ions reiterated 
these findings.

To the best of our knowledge, the only theoretical study of C$^+$ irradiation of water in the 
condensed-phase is previous work by one of us \cite{Kohanoff2008}, where projectile energies 
between 0.175 keV and 4 keV were considered, and the sample was a liquid water slab at 300 K. 
At the lowest energies complete stopping of C$^+$ was observed, with the formation of new 
chemical species like H$_2$O$_2$ and C(OH)$_2^+$.

The aim of the present study is to simulate computationally the dynamics of irradiation of 
amorphous solid water by C$^+$ ions at very low temperatures, as a prototype for the interaction 
of C$^+$ with planetary and interstellar ices. We are particularly interested in identifying which 
products emerge depending on conditions such as the kinetic energy of the carbon projectile and 
the geometrical aspects of the collision. In next Section we describe the computational approach, 
then we present the results of the simulations, and finally we summarize our conclusions.

\section{Computational Details}  

We have studied the dynamics of irradiation of water-ice clusters with low-energy singly-charged 
carbon ions at 30 K using first-principles molecular dynamics (FPMD) simulations.
The clusters have been sampled from an amorphous slab which has been supplied by 
Arasa \emph{et al.} \cite{Arasa2010}. This slab, which has been used in their studies of 
photodesorption in water-ice, has been prepared such that its structure closely resembles 
that of the compact amorphous ice obtained experimentally at 30 K.

For the purposes of this work it was deemed unnecessary to use the entire slab containing
360 water molecules. Instead, a 30-molecule sample was selected for use in 
our simulations, exactly as in our previous study with neutral C projectiles \cite{McBride2013}. 
The molecules were extracted from the uppermost portion of the slab thus representing the 
surface of the ice. Prior to the irradiation computational experiments, the system was 
quenched at 30 K for 250 fs in order to obtain initial velocities for the water molecules. 
This was achieved via a canonical (fixed temperature) FPMD simulation using 
a canonical sampling through velocity rescaling (CSVR) thermostat \cite{Bussi2007}.

Although the size of the cluster has been significantly reduced when compared with the initial 
slab, the water molecules still experience the full ice structure. One consequence of the limited 
size of this sample, however, is the increased tendency of the water molecules to evaporate during 
the lifetime of our simulations. Following the introduction of the energetic projectile the 
temperature of the system rises enormously. Impact by an 11 eV carbon ion will cause the temperature 
of the system to rise to approximately 85000 K while an average energy of roughly 0.25 eV is imparted 
to each water molecule. This effect would surely be reduced through use of an enlarged ice sample
or a heat sink at the boundary of the cluster. It is, however, a genuine feature of the irradiation 
process, that the sample experiences a huge local temperature increase that favors chemical
reactivity in the time scales studied in the present work.

All simulations were carried out using the ab initio module QUICKSTEP of the CP2K package 
\cite{VandeVondele2005}. The electronic structure was computed at the all-electron level, within 
density functional theory (DFT) using the Gaussian and augmented plane waves method (GAPW). Here 
the Kohn-Sham orbitals are expanded in a Gaussian basis set while the electrostatic (Hartree) 
potential is computed by expressing the electronic density in plane waves (PW) augmented with 
atom-centered Gaussian functions. The latter are especially important within the framework of 
all-electron calculations because of the core orbitals, which would require PW with a very large 
energy. The augmentation approach allows for a significant reduction of the PW cutoff, to levels
similar to those required in pseudopotential calculations.

\begin{figure}[t]
 \noindent\makebox[\textwidth]{
 \includegraphics[width=150pt]{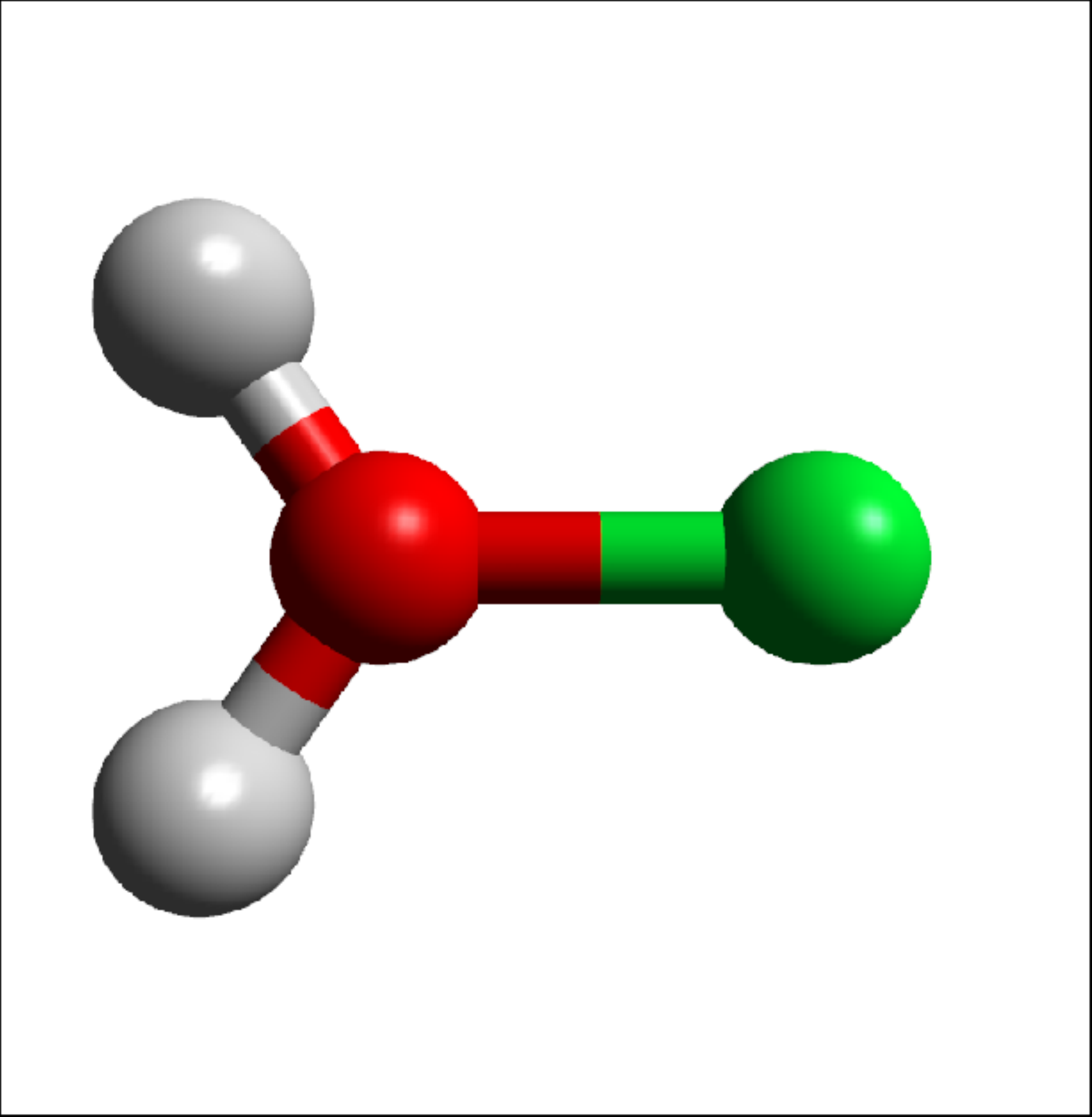}
 }
 \caption{Optimized geometry for the C$^+$-water complex. The C-O distance is given in 
 \ref{tab:1} for various theory levels.}
 \label{fig:opt_coh2}
\end{figure}

After testing the quality of various approximations to exchange and correlation, we chose to 
use the hybrid PBE0 functional, which includes a Hartree-Fock exchange contribution combined 
with the Perdew-Burke-Ernzerhof (PBE) semi-local exchange-correlation term \cite{Perdew1996}. 
The 6-311G** basis set was selected as a good compromise between accuracy and computational 
efficiency. With this choice, basis set superposition errors are smaller than 3 \%. The PBE0 
functional was chosen based on the comparison of binding energies,

\begin{equation} \label{eq:energy}
 E_b = -[E(COH_2^+)-E(C^+)-E(H_2O)]
\end{equation}
for the C$^+$-water complex, see \ref{fig:opt_coh2}, against quantum chemical calculations 
at the MP2 and CCSD(T) level. The results of these calculations are reported in \ref{tab:1}. 
These computations were carried out using the NWChem package \cite{NWChem}. The geometry of 
the complex was confirmed following extensive optimization calculations performed at the 
MP2/aug-cc-pVTZ level.

\begin{table}[t]
\noindent\makebox[\textwidth]{
\begin{tabular}{ l l c c }
  \hline
  Theory Level & Basis set & Binding energy (eV) & C-O bond length (\AA) \\
  \hline
  HF      & 6-311G**     & 3.40 & 1.39\\
  PBE     & TZVP         & 4.60 & 1.40\\
  PBE0    & TZVP         & 4.38 & 1.39\\
  MP2     & aug-cc-pVTZ  & 4.02 & 1.37\\
  CCSD(T) & aug-cc-pVTZ  & 3.96 & 1.38\\
  \hline
\end{tabular}
}
\caption{Calculated binding energies and $\mathrm{COH_2^+}$ bond lengths
for the C$^+$-water complex.}
\label{tab:1}
\end{table}

The calculated binding energies depend on the theory level, with Hartree-Fock and PBE at 
the two extremes, respectively under- and over-binding. The binding energies produced by 
PBE0, MP2 and CCSD(T) are all approximately within 10 \% of each other -- PBE0 is within 
10.6 \% of CCSD(T). 

Notice the significant binding energy difference with respect to the neutral complex, 
examined in our previous work \cite{McBride2013}, where the theory level was much more 
relevant. Analysis of the electronic density reveals that the carbon ion actually makes 
a chemical bond with the water molecule. The C-O distance of 1.39 \AA~is significantly 
longer than the 1.13 \AA~in the CO molecule, but this is not surprising since the latter
is a triple bond while in COH$_2^+$ we have a situation closer to a single bond. This is 
similar to the neutral COH$_2$ species which we previously reported \cite{McBride2013}, 
where the C-O bond was 0.05 \AA~ longer due to the additional electron. 
A consequence of the formation of this chemical bond is that the O-H bonds in the water 
molecule are weakened -- elongating from 0.96 to 0.99 \AA~--, thus facilitating proton 
transfer from COH$_2^+$ to the neighboring water molecules in the condensed phase.

In respect of hydrogen bonds, it is well-documented that PBE0, due to the dominance 
of electrostatics, renders the geometries and energetics quite well in comparison to 
higher-level calculations and dispersion-corrected functionals \cite{Thanthiriwatte2011}. 
As a check, equilibration of the cluster was repeated with the inclusion of Grimme's
semi-empirical van der Waals interaction \cite{grimme}. Comparison of the O-O distances 
for each cluster indicates that no significant changes result from the addition of these 
dispersion forces. To better assess the effect of the missing dispersion in PBE0, a 
trajectory in which COH$\rad$ forms was repeated with the inclusion of the van der Waals 
interaction. There was no major change to the outcome of this trajectory, which once again
produced COH$\rad$ (see supporting information).

\begin{figure}[t]
 \noindent\makebox[\textwidth]{
 \includegraphics[width=150pt]{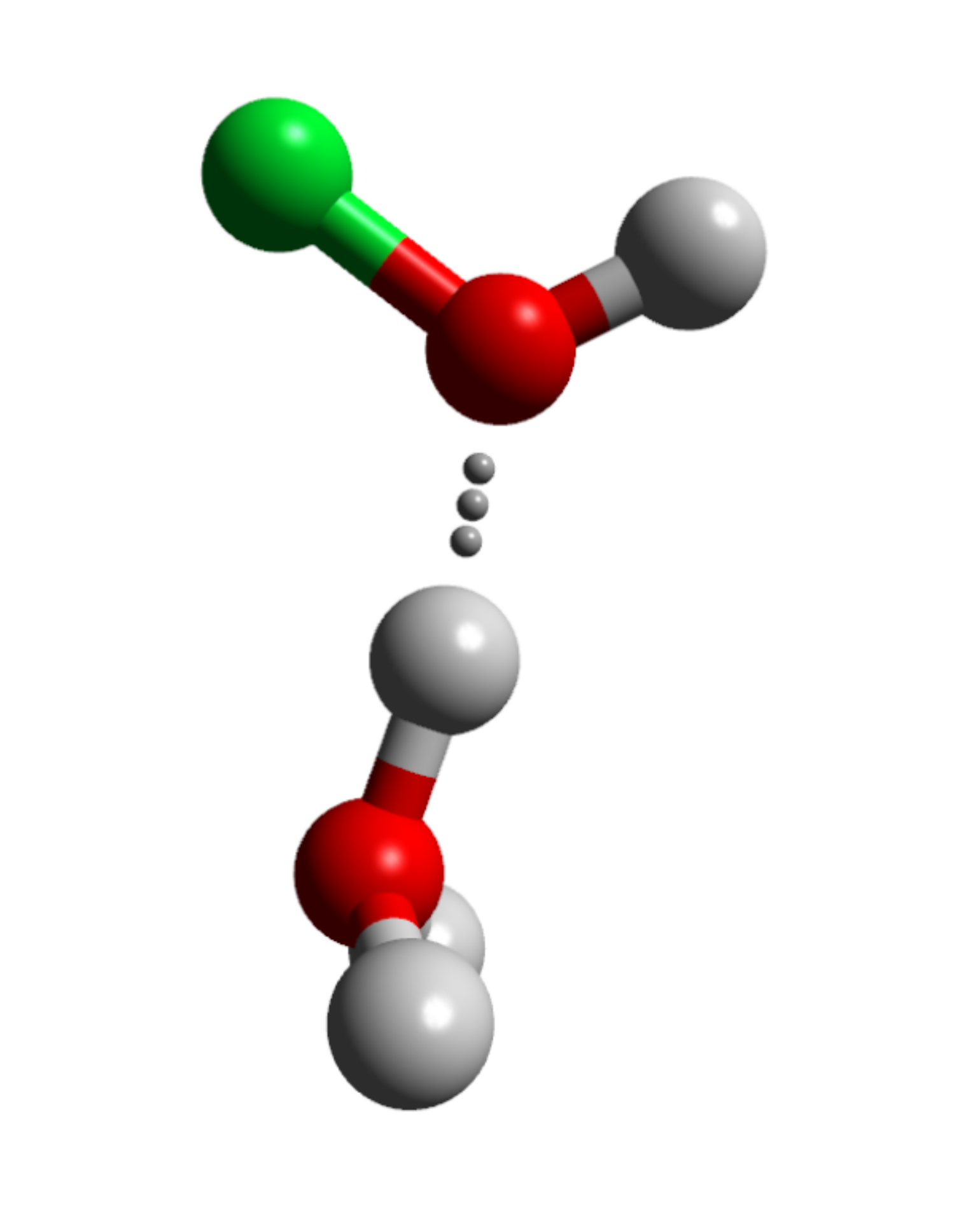}
 }
 \caption{Optimized geometry for the COH$\cdots$H$_3$O$^+$ complex. Dots indicate the hydrogen bond,
with an O$\cdots$H distance of 1.47 \AA~at the PBE0/TZVP level (1.50 \AA~at the MP2/aug-cc-pVTZ level).
The binding energy is approximately 0.7 eV.}
 \label{fig:coh_h3o+_opt}
\end{figure}

As a further test of the PBE0 functional we have compared the binding energy and hydrogen 
bond length for COH$\rad$-hydronium complex (see \ref{fig:coh_h3o+_opt}) at the PBE0/TZVP 
and MP2/aug-cc-pVTZ level. This entity appears in practically all the FPMD simulations.
The calculated binding energies are 0.69 eV (PBE0) and 0.73 eV (MP2), while the bond 
lengths measure 1.47 {\AA} and 1.50 {\AA}. We now find that the disparity between the 
levels of theory which had been observed earlier (see \ref{tab:1}) is greatly reduced. 
It is also apparent from our calculations that there is no barrier to the formation of 
COH$\rad$ in the gas-phase reaction

\begin{equation}\label{eq:c++2h2o}
 C^+ + 2H_{2}O \rightarrow COH\rad + H_3O^+ \qquad .
\end{equation}
We believe this level of agreement is sufficient to describe, with 
reasonable accuracy, collisions where the C$^+$ ion binds to a water molecule.
Although MP2 generally outperforms PBE0, the much higher computational expense incurred 
through its use is, in our opinion, not justifiable, while this type of simulations are
simply not viable at the CCSD(T) level of theory.

The configuration of water molecules obtained following the completion of the canonical 
FPMD run was used as a starting point for each of the subsequent simulations. Initial 
kinetic energies of 1.7 eV and 11 eV were chosen for the carbon projectile. Those two
energies were selected as representative of two distinct situations. At 11 eV the 
projectile has the energy required to dissociate a water molecule (5.1 eV), while
at 1.7 eV water dissociation must proceed through other, more complex channels.
For each of these energies 15 microcanonical (total energy of the system is fixed) runs 
were completed where, in each case, the carbon projectile had a different starting position 
and/or direction of approach. We also looked at higher-energy projectile, but these pass
right through the cluster without stopping, thus not generating new chemical species.
Clearly, 30 simulations are not statistically significant in order to extract information 
about probabilities and cross sections. However, the goal of the present work is to identify 
collision channels rather than to estimate cross sections.

\section{Results}

Reported here are the products that result from the reactions following relatively 
low-energy collisions between the carbon ion and water-ice clusters. Our simulations 
have produced the following carbon bearing species:
\\

Formyl and isoformyl radicals:
\begin{equation}\label{eq:HCO}
 C^+ + nH_2O \rightarrow HCO\rad + H_3O^+ + (n-2)H_2O
\end{equation}

\begin{equation}\label{eq:COH}
 C^+ + nH_2O \rightarrow COH\rad + H_3O^+ + (n-2)H_2O
\end{equation}

carbon monoxide:
\begin{equation}\label{eq:CO}
 C^+ + nH_2O \rightarrow CO + H\rad + H_3O^+ + (n-2)H_2O
\end{equation}

and the dihydroxymethyl radical:
\begin{equation}\label{eq:CHOH2}
 C^+ + nH_2O \rightarrow CH(OH){\rad}_2 + H_3O^+ + (n-3)H_2O
\end{equation}

where $n$ is the number of water molecules, in this case $n$ = 30.
\\\\
In all cases, in the aftermaths of the collision, the positive charge 
of the system is held by a proton that forms a hydronium ion in the 
condensed-phase environment.

\subsection{Isoformyl Formation}

\ref{eq:COH} describes the production of the isoformyl radical (COH$\rad$). 
This molecule appears as the final product in a third of our simulations. We find that 
the formation of COH$\rad$ radicals is favored to a larger extent at lower energies.

\begin{figure}[ht]
\centering
\subfloat[20 fs] {\includegraphics[width=110pt]{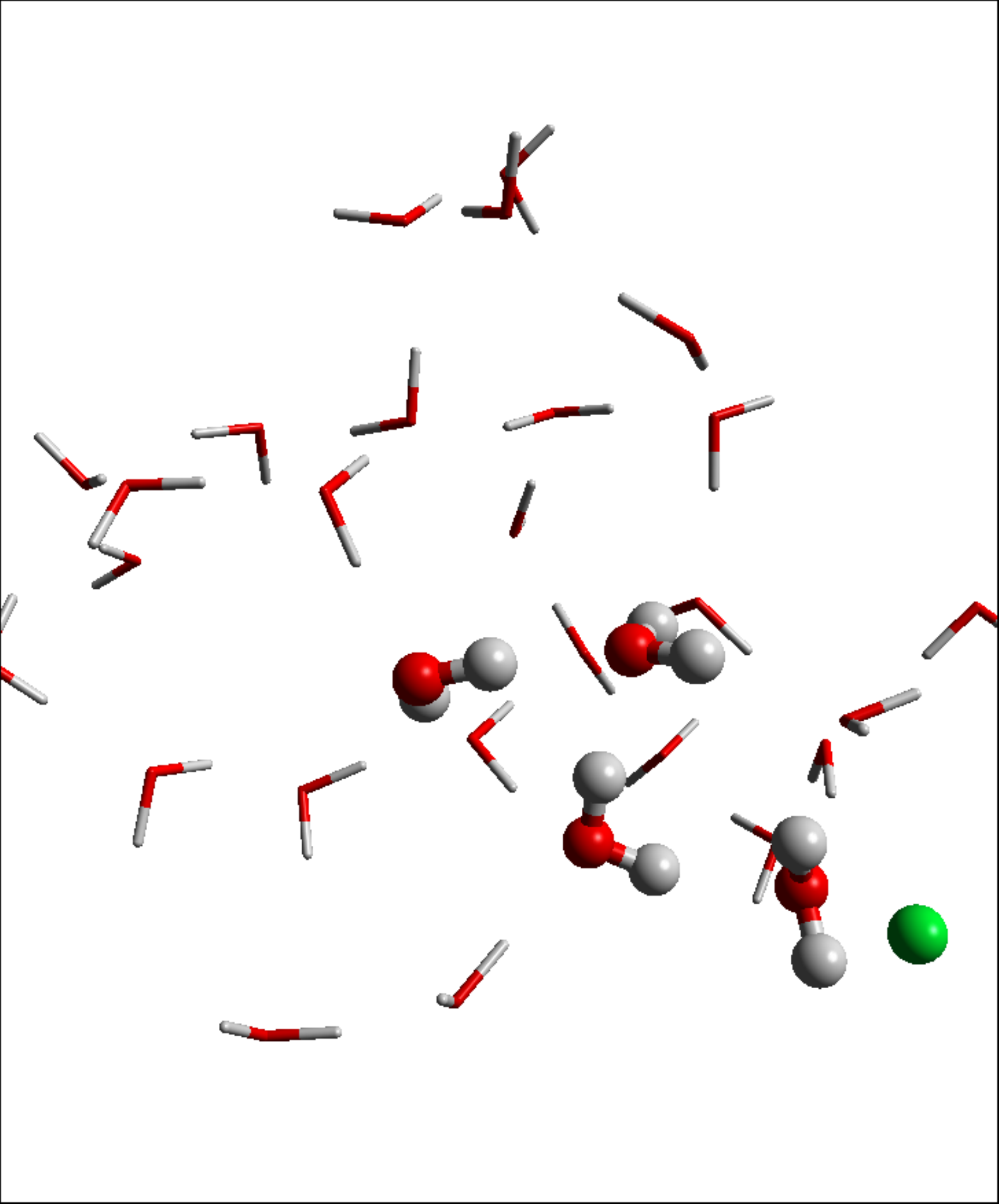} \label{fig:coh1}}
\subfloat[43 fs] {\includegraphics[width=110pt]{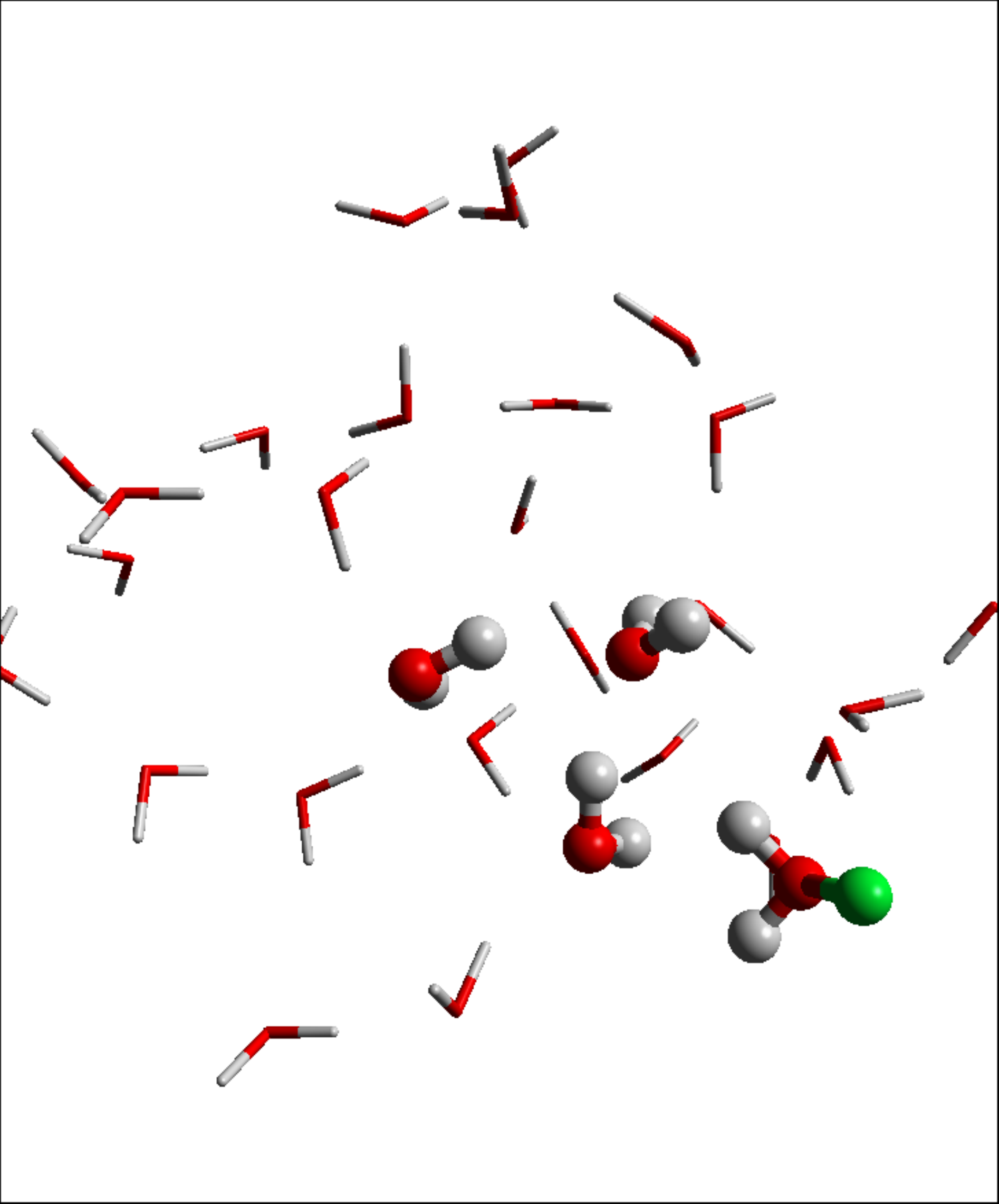} \label{fig:coh2}}
\subfloat[120 fs]{\includegraphics[width=110pt]{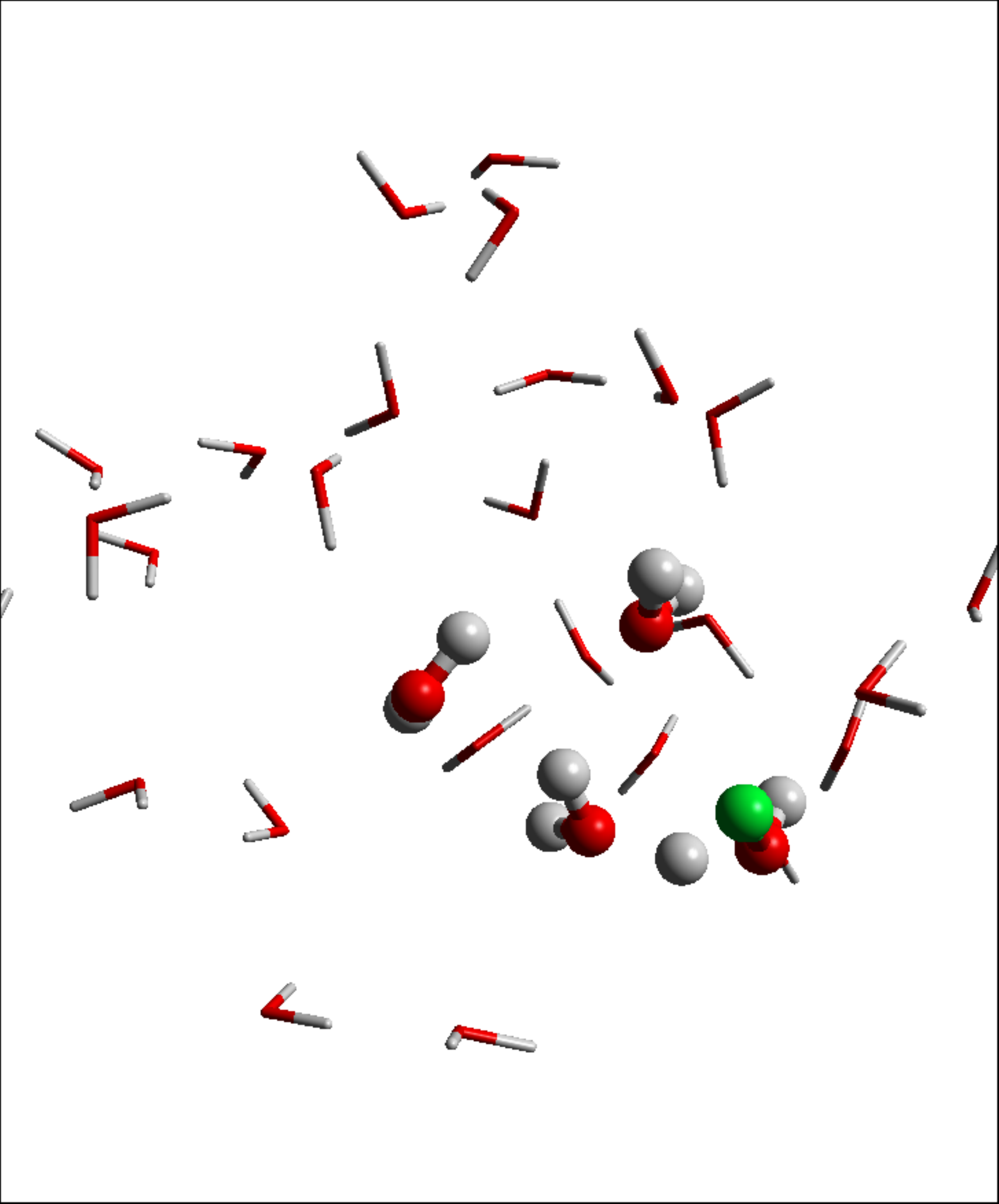} \label{fig:coh3}}\\
\subfloat[127 fs]{\includegraphics[width=110pt]{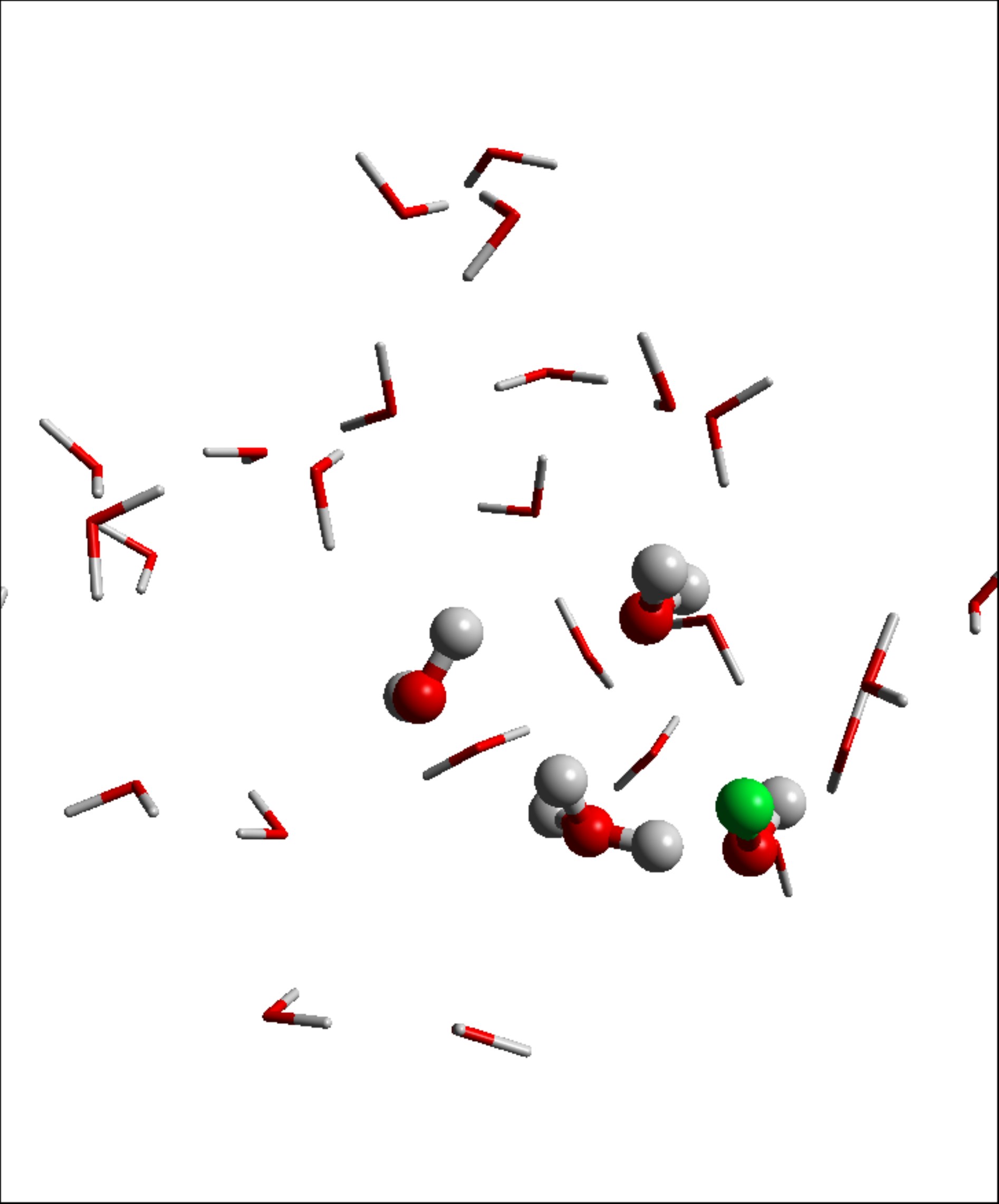} \label{fig:coh4}}
\subfloat[187 fs]{\includegraphics[width=110pt]{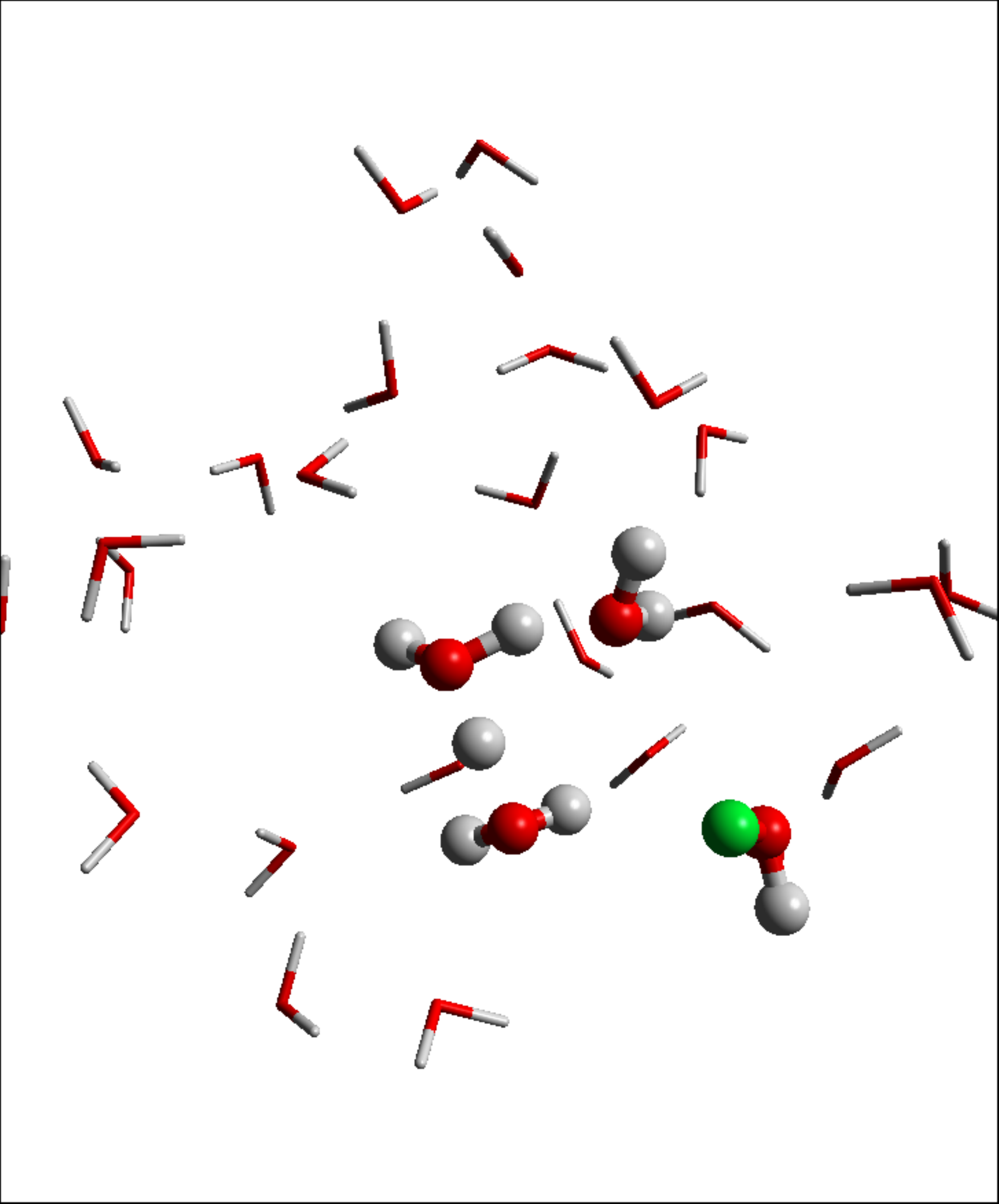} \label{fig:coh5}}
\subfloat[198 fs]{\includegraphics[width=110pt]{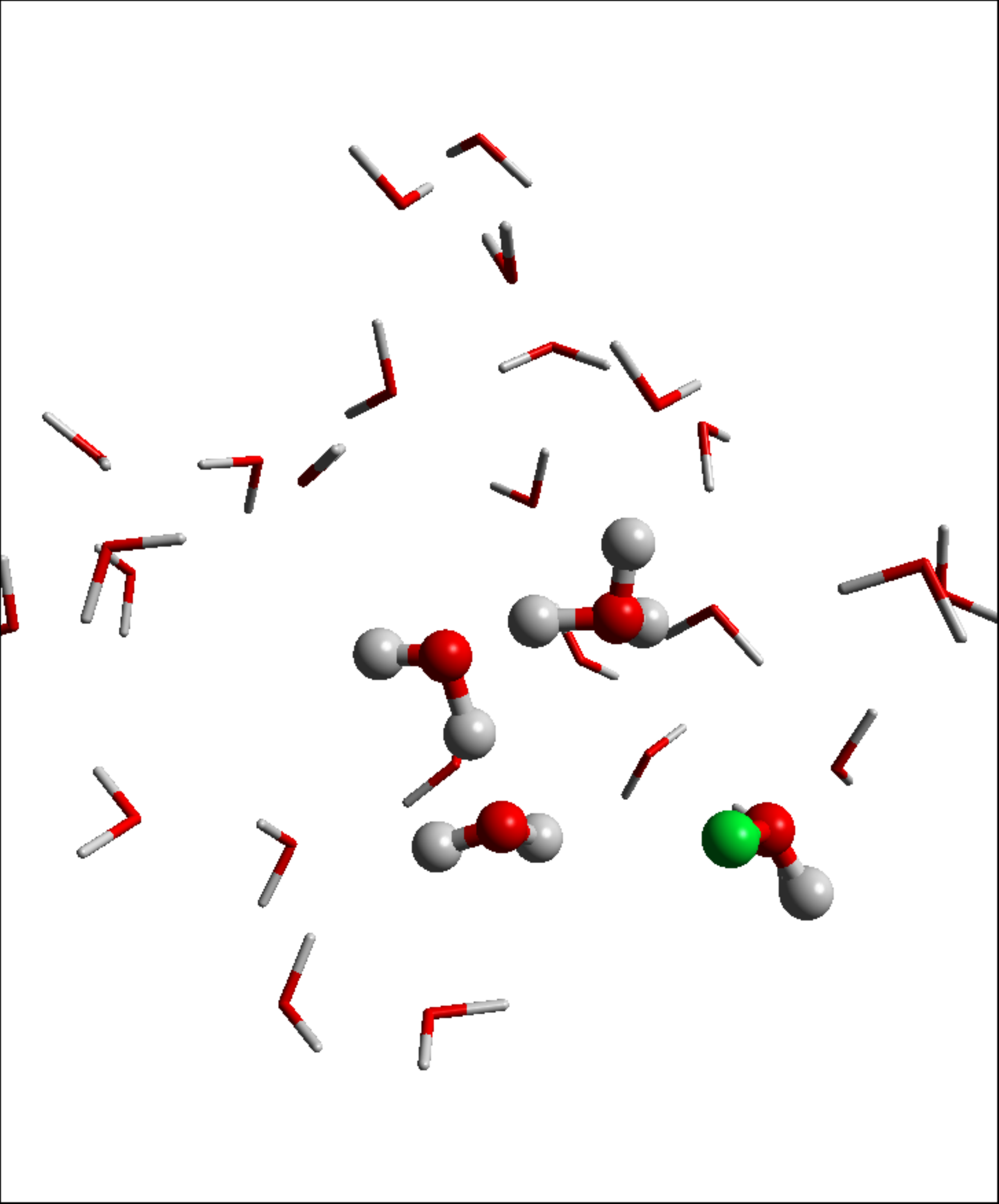} \label{fig:coh6}}
\caption{A selection of frames from a trajectory in which the isoformyl radical 
(COH\protect\rad) forms. The atoms/molecules which participate directly in the 
reaction have been represented by \emph{spheres}. The remaining water molecules 
are shown as \emph{tubes}.}
\label{fig:1}
\end{figure}

The reaction mechanism for the formation of COH$\rad$ has been identified. The sequence 
of steps for such a trajectory are depicted in \ref{fig:1}.

As the carbon ion approaches the cluster it makes a bond with the oxygen of a 
water molecule upon impact \emph{(b)}. It has been found that at this point the 
reaction may proceed in either one of two ways. Depending on the energy of the 
carbon ion and its angle of impact, the COH$\rad$ radical may be produced directly 
or through an intermediate. In the direct, ``knock-out'', mechanism the carbon 
ion simply ejects a proton from the water molecule, which then goes on to form a 
hydronium ion, H$_3$O$^+$, in the cluster. This is a relatively rapid process and 
occurs when the C$^+$ projectile is more energetic. Shown in \ref{fig:1}\emph{(b)} 
is the formation of an intermediate, COH$_2^+$. When compared with the ``knock-out'' 
mechanism (not shown), the carbon ion is bonded to the water molecule for significantly 
longer (about 80 fs) before a proton detaches.

As the reaction proceeds we see that as the carbon ion forms a bond with the water
molecule, it weakens one of its OH bonds eventually leading to the detachment of a 
proton \emph{(c)}. This proton subsequently joins a nearby water to form a positively 
charged hydronium \emph{(d)}, leaving the neutral COH$\rad$ radical. Following this 
it was possible to observe proton transfer from hydronium to neighboring water 
molecules within the cluster, \emph{(e)} and \emph{(f)}. This continues until a stable 
arrangement is found, after dissipating the necessary amount of kinetic energy. 
H$_3$O$^+$ is a dynamical species at room temperature, with the proton diffusing via 
the Grotthuss mechanism \cite{Grotthuss1806, Agmon1995}. However, at the low 
temperatures considered here, the crucial step of reorientation of water molecules 
to create the appropriate environment for proton transfer is kinetically hindered. 
Therefore, the proton transfer stops after a few steps. The kinetic energy of the 
projectile is transferred to the water molecules, which occasionally evaporate from 
the cluster. At this point the simulation becomes unrealistic because a significant 
part of this energy should be dissipated through the ice environment, which is absent
in the cluster geometry. Hence, we stop the simulation at this stage.  

\subsection{Formyl Formation}

In agreement with the literature pertaining to gas phase MD calculations, we have 
found that production of the formyl (HCO) species does not occur as readily as isoformyl. 
However, it has been observed that under the conditions of our simulations HCO$\rad$ 
creation is more common here than in the gas phase. Notice, however, that in the condensed
phase we observe neutral radicals instead of cations (as in the gas phase). This may be due
to the moderating presence of the ice environment, which solvates and screens the
positively charged protons. In the gas phase dissociation proceeds by losing 
a hydrogen atom instead of a proton.

\ref{eq:HCO} describes the production of the HCO$\rad$ radical. The molecule 
was found as the final product in five of the trajectories. A tendency to favor low-energy 
trajectories was also expressed here.

\begin{figure}[ht]
\centering
\subfloat[0 fs]  {\includegraphics[width=110pt]{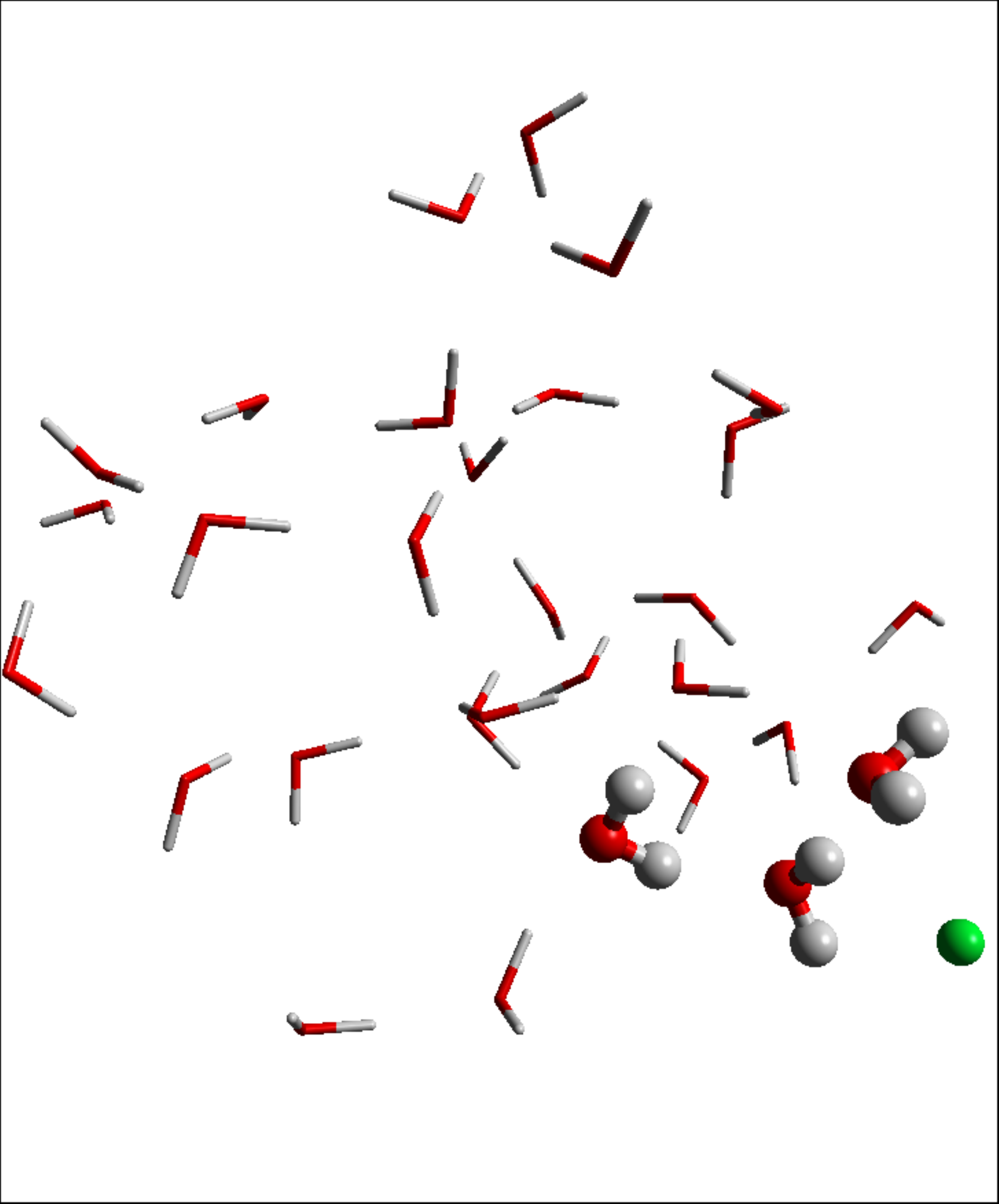} \label{fig:hco1}}
\subfloat[30 fs] {\includegraphics[width=110pt]{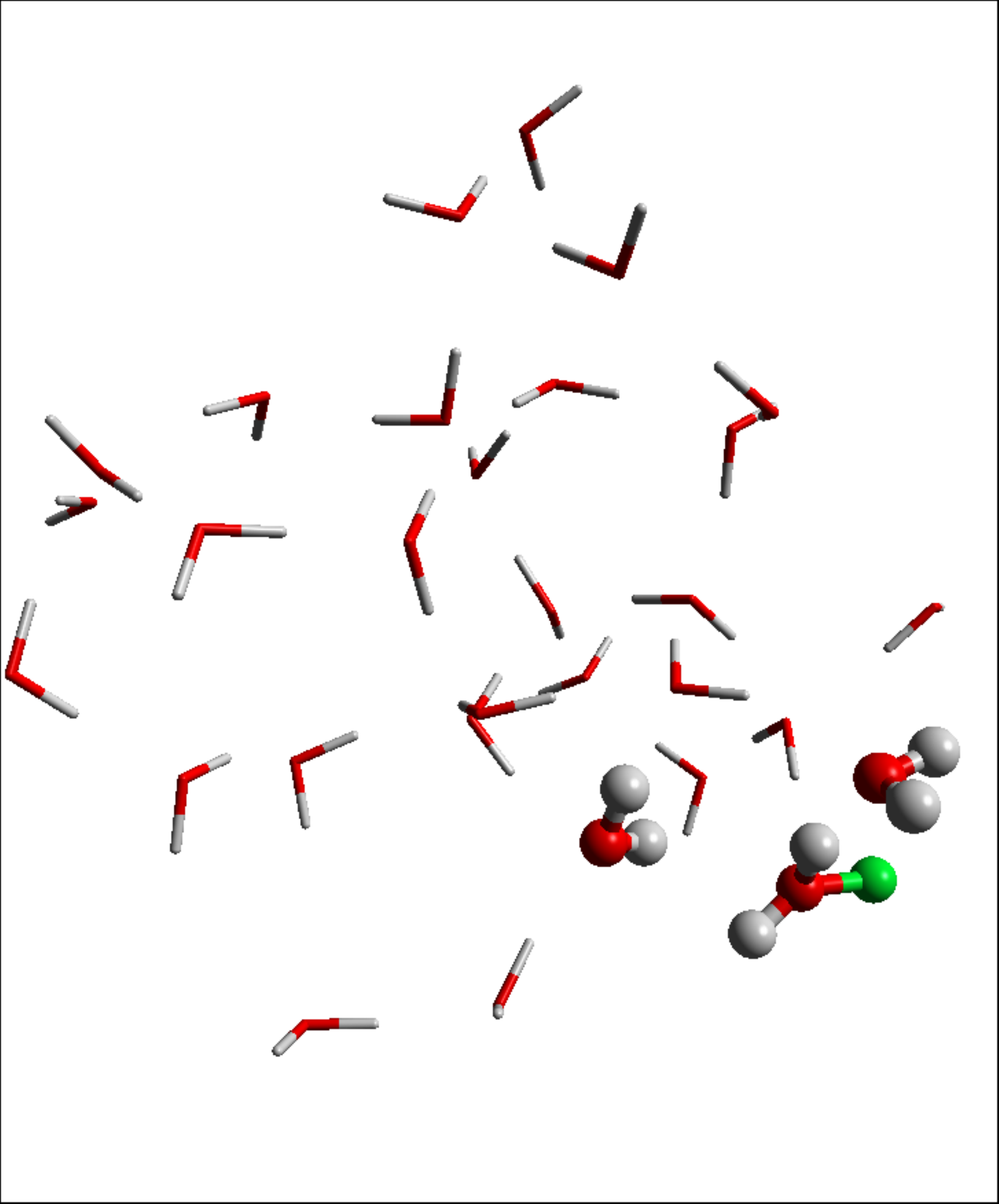} \label{fig:hco2}}
\subfloat[57 fs] {\includegraphics[width=110pt]{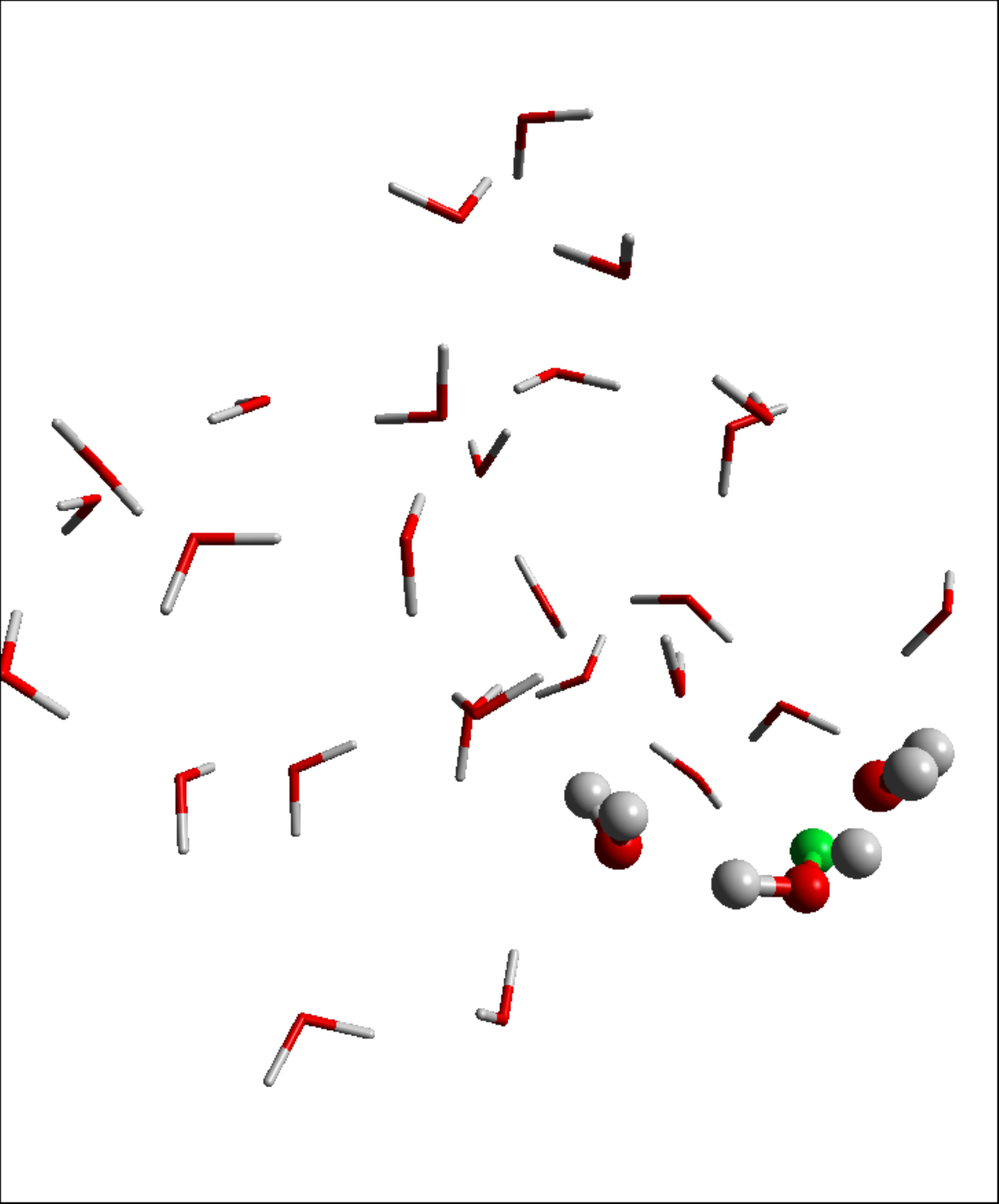} \label{fig:hco3}}\\
\subfloat[72 fs] {\includegraphics[width=110pt]{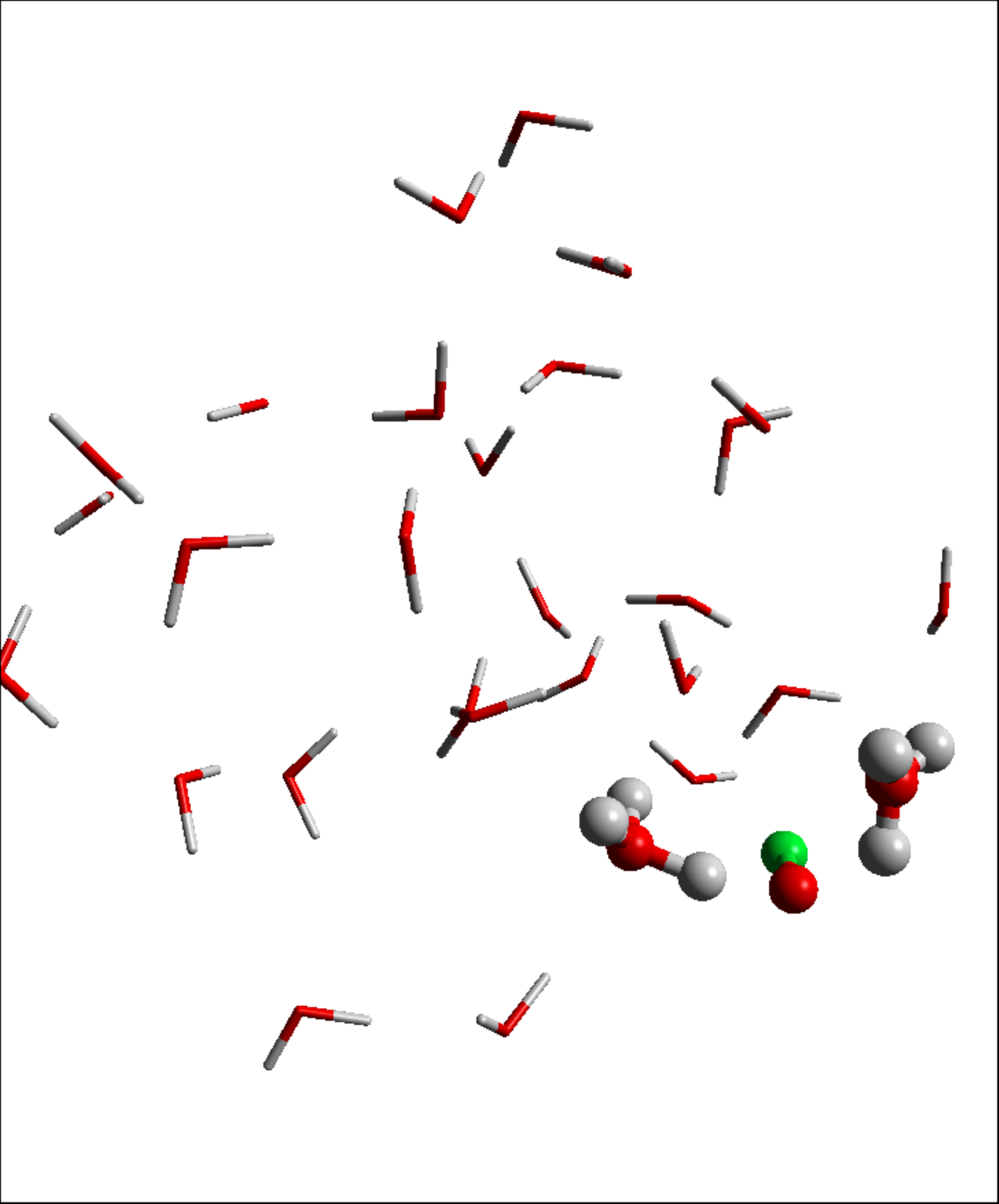} \label{fig:hco4}}
\subfloat[104 fs]{\includegraphics[width=110pt]{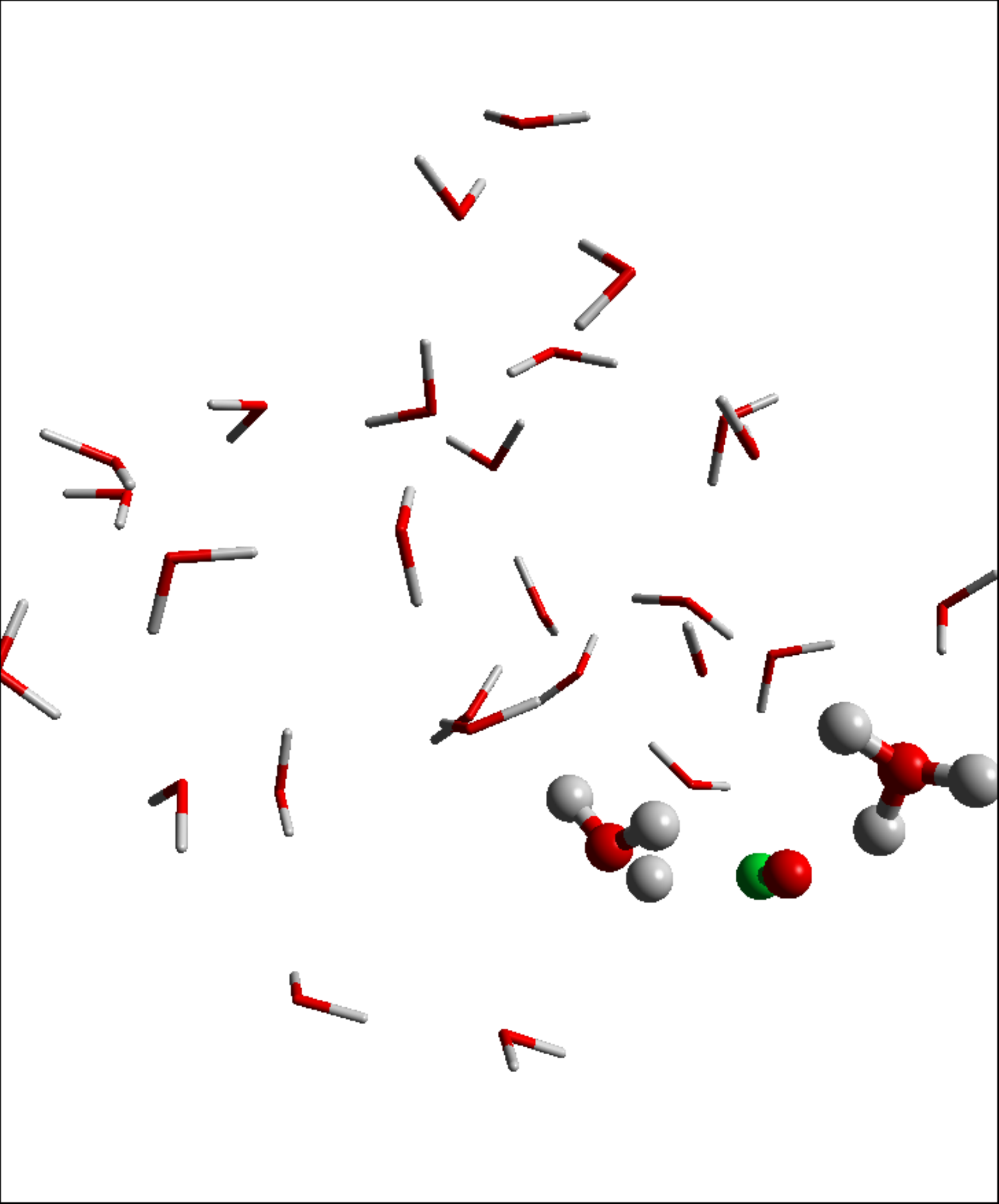} \label{fig:hco5}}
\subfloat[114 fs]{\includegraphics[width=110pt]{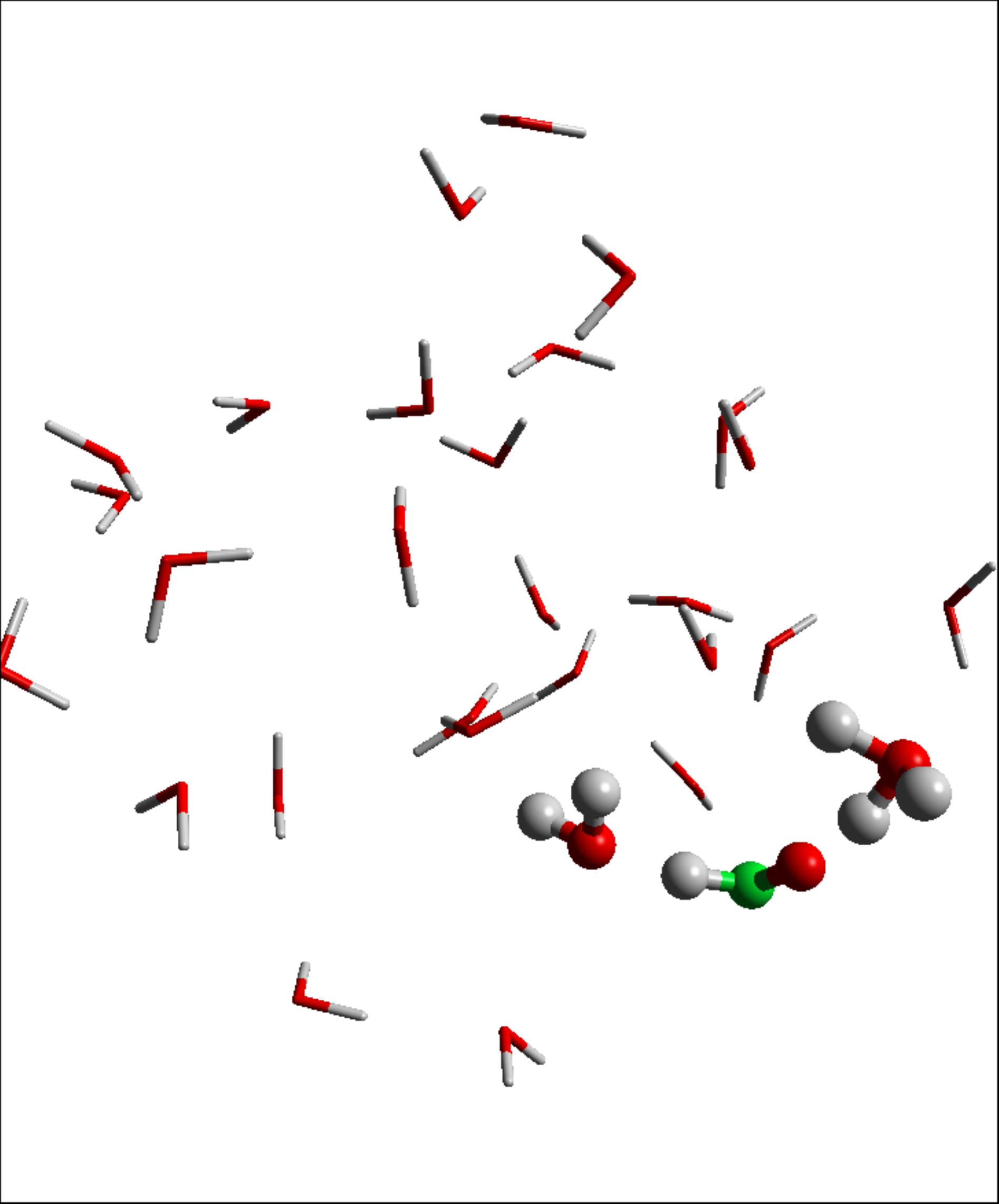} \label{fig:hco6}}
\caption{A selection of frames from a trajectory in which the formyl radical 
(HCO\protect\rad) forms. The atoms/molecules which participate directly in the 
reaction have been represented by \emph{spheres}. The remaining water molecules 
are shown as \emph{tubes}.}
\label{fig:2}
\end{figure}

\ref{fig:2} shows the most common route to the formation of HCO$\rad$. 
The process begins similarly to that seen in the formation of COH$\rad$. In 
this case two protons detach following the initial impact of the carbon 
ion \emph{(d)}. A CO$^-$ radical is left temporarily surrounded by a pair 
of hydronium molecules. Subsequently, a proton from one of the hydronium 
molecules rebinds to the carbon. The neutral HCO$\rad$ radical is thus
produced, leaving one remaining positively charged hydronium, H$_3$O$^+$.

\begin{figure}[ht]
\centering
\subfloat[35 fs] {\includegraphics[width=110pt]{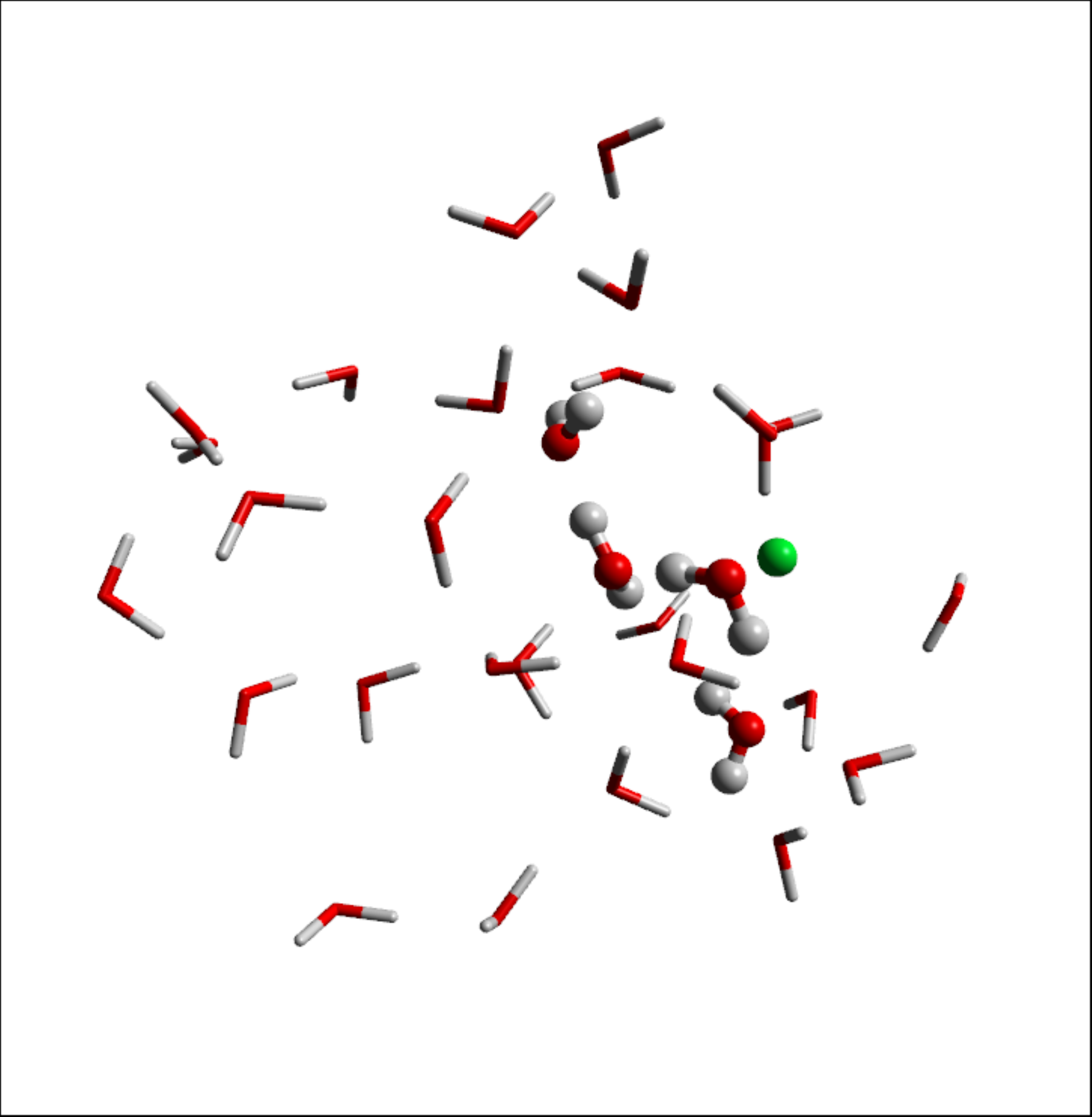} \label{fig:2hco1}}
\subfloat[40 fs] {\includegraphics[width=110pt]{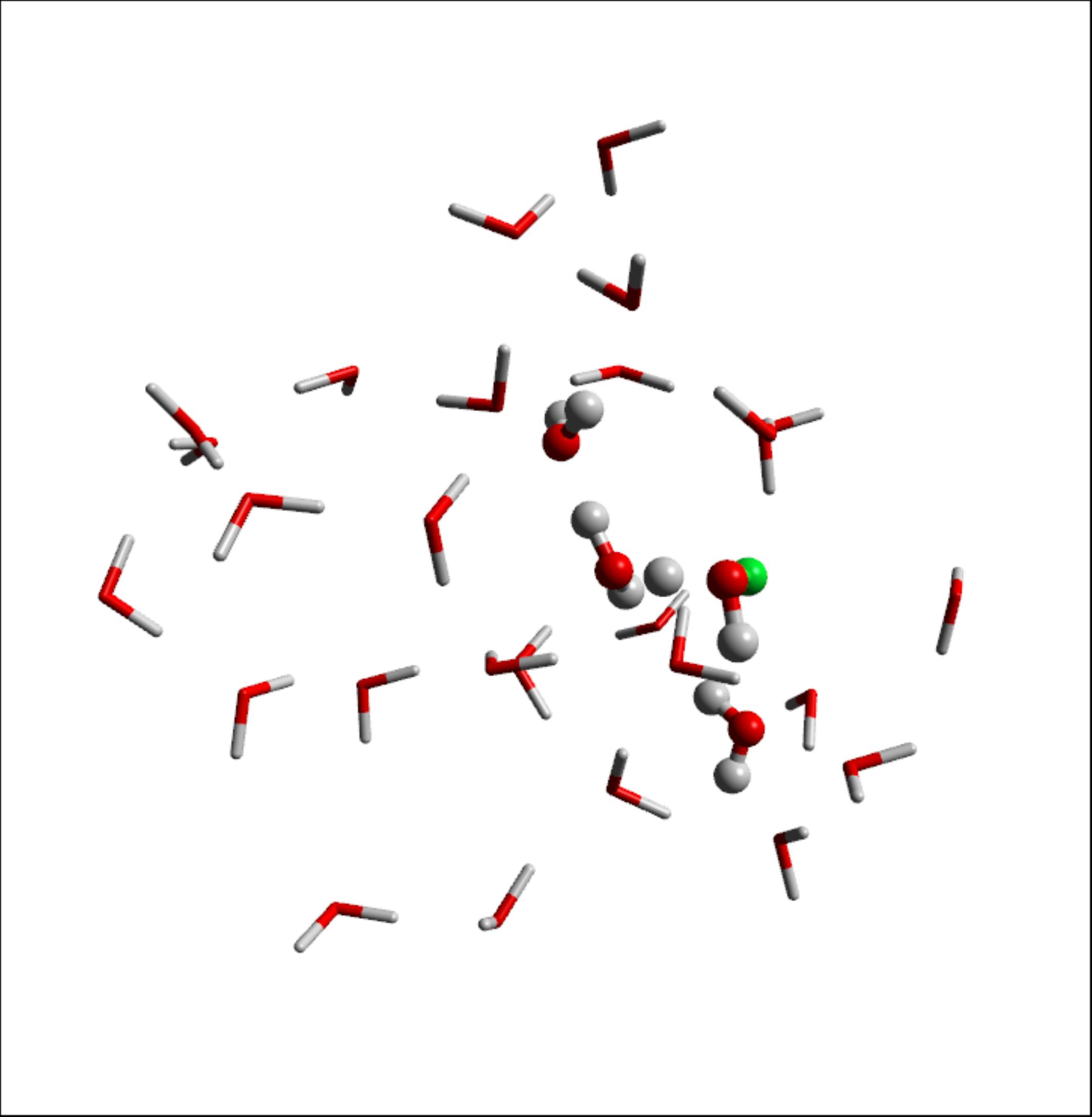} \label{fig:2hco2}}
\subfloat[50 fs] {\includegraphics[width=110pt]{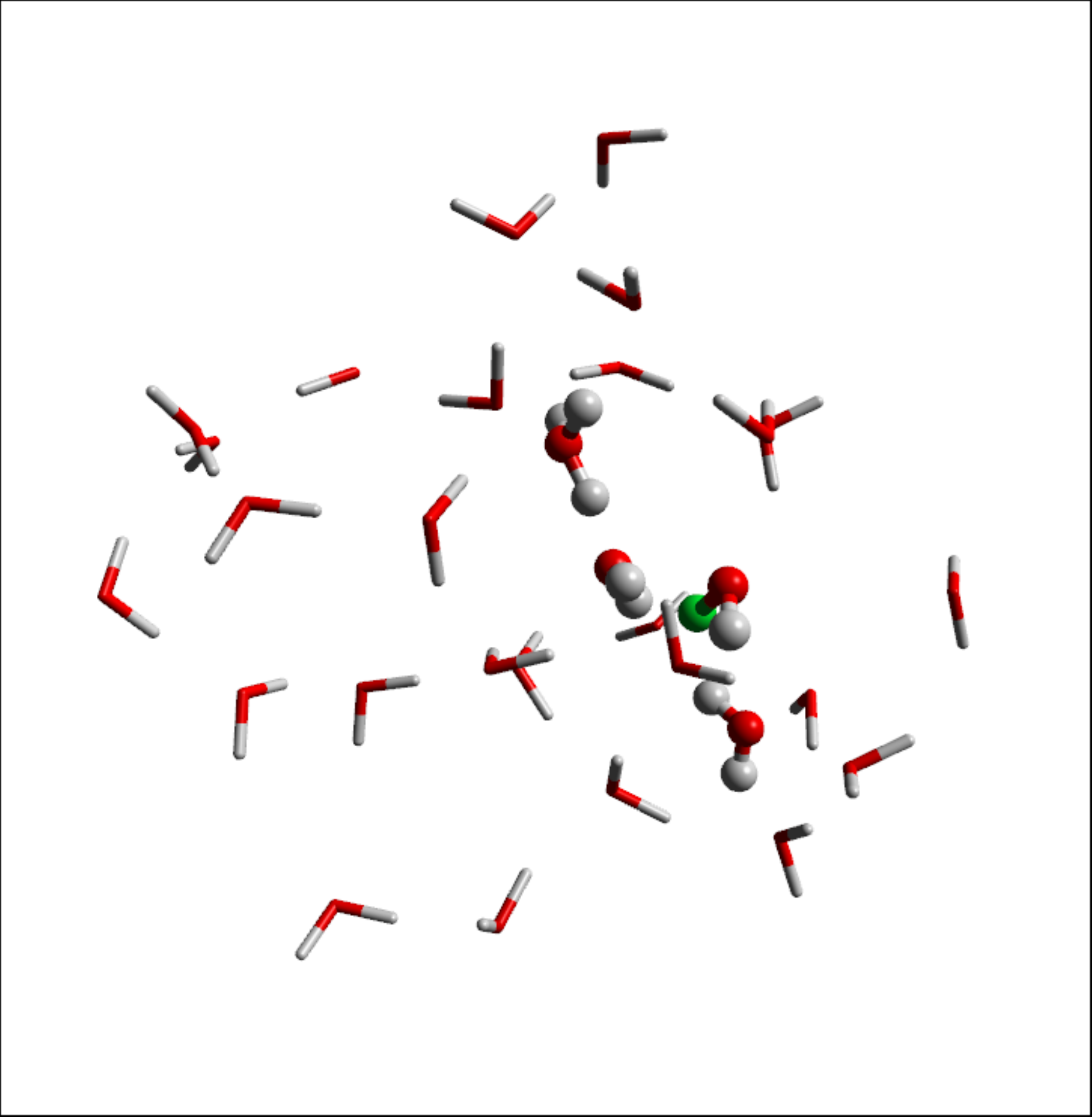} \label{fig:2hco3}}\\
\subfloat[65 fs] {\includegraphics[width=110pt]{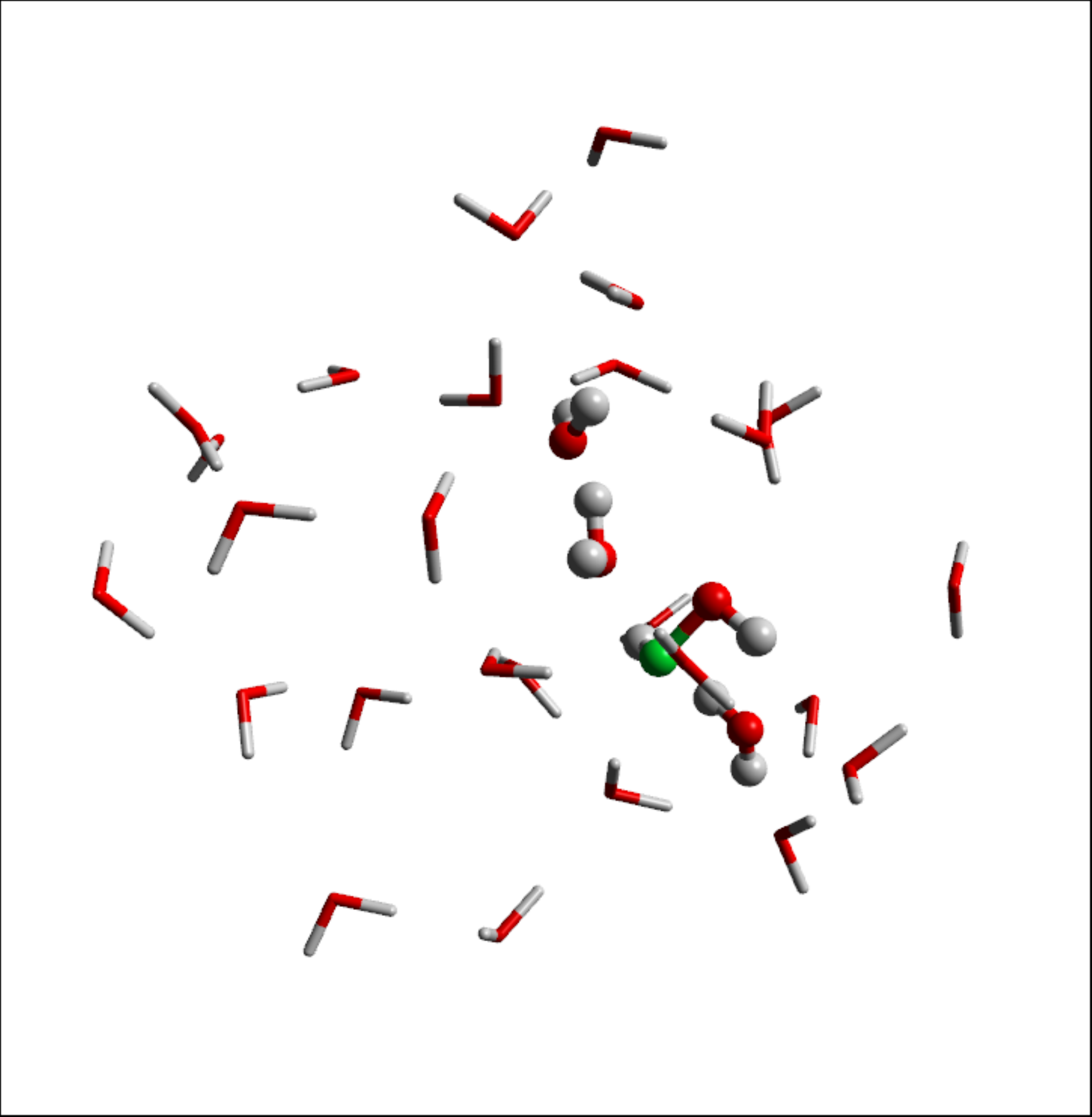} \label{fig:2hco4}}
\subfloat[110 fs]{\includegraphics[width=110pt]{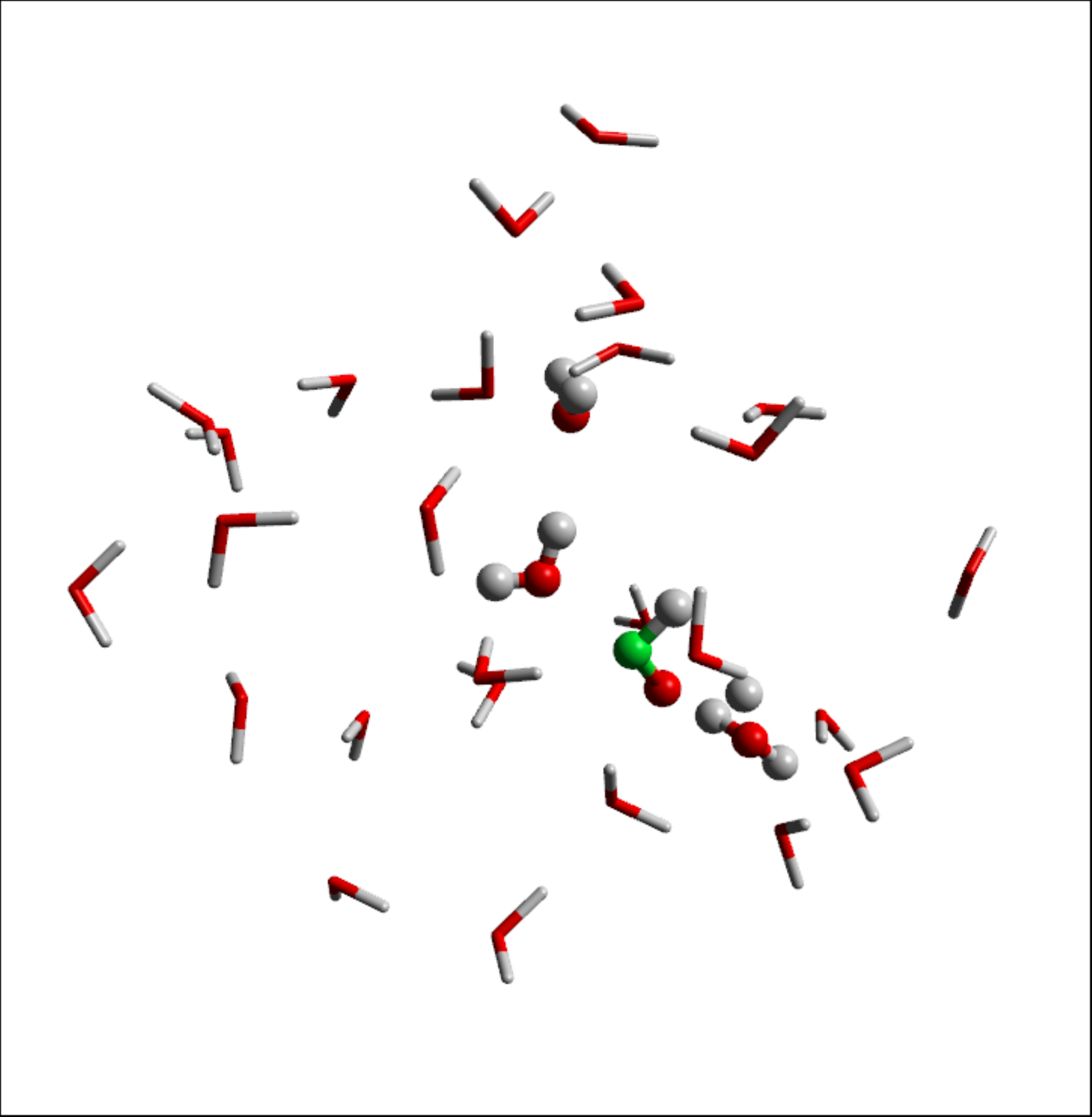} \label{fig:2hco5}}
\subfloat[125 fs]{\includegraphics[width=110pt]{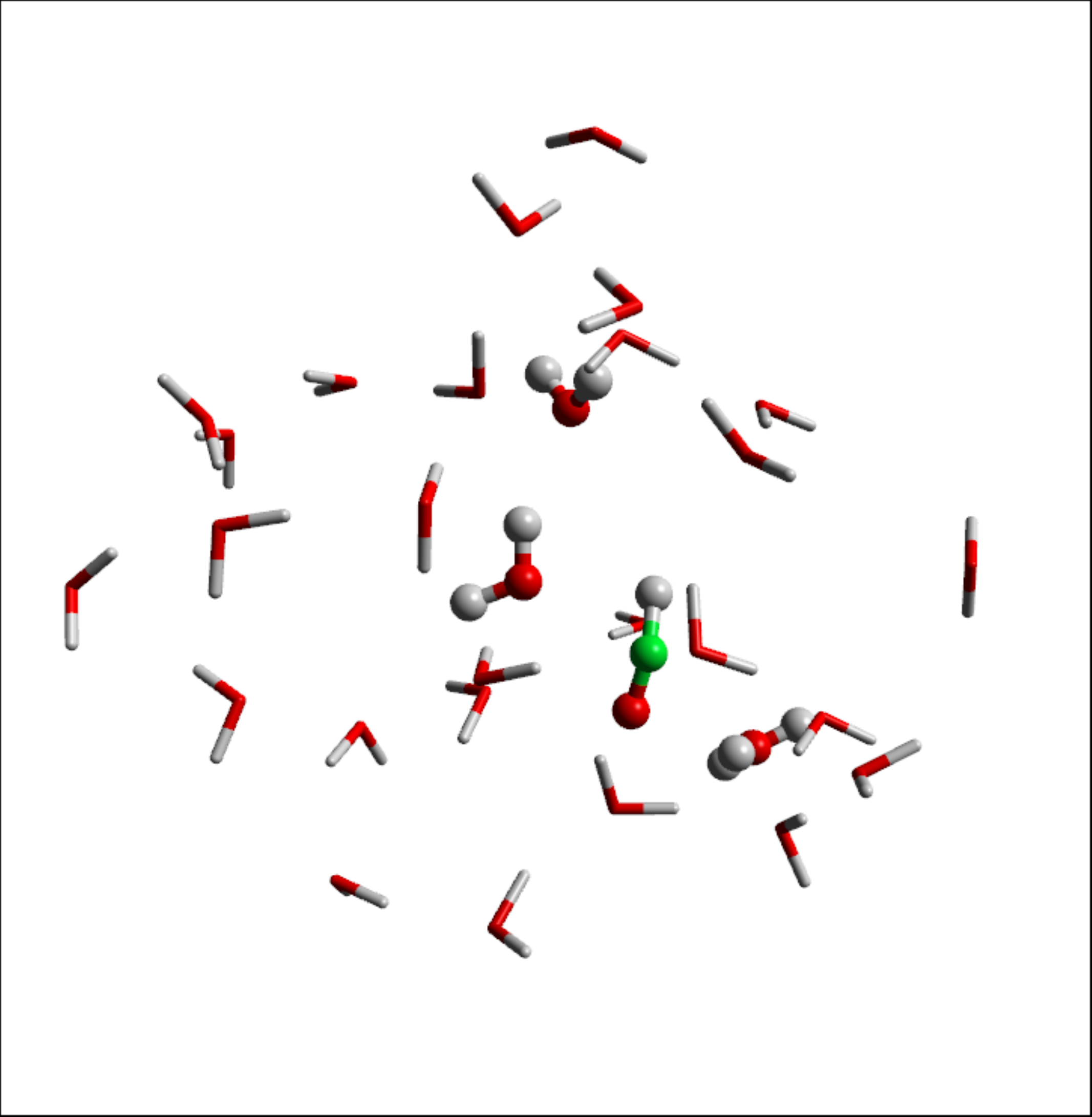} \label{fig:2hco6}}
\caption{A selection of frames from a trajectory in which the formyl radical 
(HCO\protect\rad) forms via HCOH$^+$. The atoms/molecules which participate directly in the 
reaction have been represented by \emph{spheres}. The remaining water molecules 
are shown as \emph{tubes}.}
\label{fig:2a}
\end{figure}

On one occasion we also observed HCO$\rad$ production via formation of an 
HCOH$^+$ intermediate (see \ref{fig:2a}), as described by Ishikawa for the gas
phase \cite{Ishikawa2001}. Following the initial formation of the COH$\rad$ radical 
\emph{(b)} the molecule repositions itself allowing for the attachment of a 
proton from a neighboring hydronium species to the carbon atom \emph{(d)}. 
A relatively long period of time ($\sim$ 50 fs) now elapses in which the 
hydroxymethylene ion reorients enabling the transfer of the oxygen-bonded 
proton to a water molecule \emph{(e)}.

\subsection{Carbon Monoxide Formation}

\ref{eq:CO} describes the production of carbon monoxide, CO, which formed 
in 40 \% of our simulations. Trajectories in which the carbon ion had the higher 
kinetic energy of 11 eV were favored for CO production by approximately a 
factor of 2 with respect to those at the lower energy. The mechanism for its formation 
is not unique. CO is created when a hydrogen atom detaches from the newly created COH$\rad$ 
or HCO$\rad$ radicals, usually COH$\rad$. This (neutral) hydrogen atom will often evaporate 
from the ice cluster, as its interaction with water molecules is very weak. In fact, our
calculations show that the H--H$_2$O binding energy is of the order of a few meV.
In a larger sample of ice molecules, this hydrogen atom may act as a secondary projectile.
Such low-energy hydrogens generally react with other radicals saturating 
their valence, as shown in the following reactions:
\\
\begin{equation}\label{eq:H2COH}
 HCOH + H \rightarrow H_2COH
\end{equation}

\begin{equation}\label{eq:H3COH}
 H_2COH + H \rightarrow H_3COH \textrm{ (methanol)}
\end{equation}
that are of crucial relevance in the formation of methanol on icy grains. We will look at 
these reactions in a forthcoming paper.

\subsection{Dihydroxymethyl Formation}

\begin{figure}[ht]
\centering
\subfloat[73 fs] {\includegraphics[width=110pt]{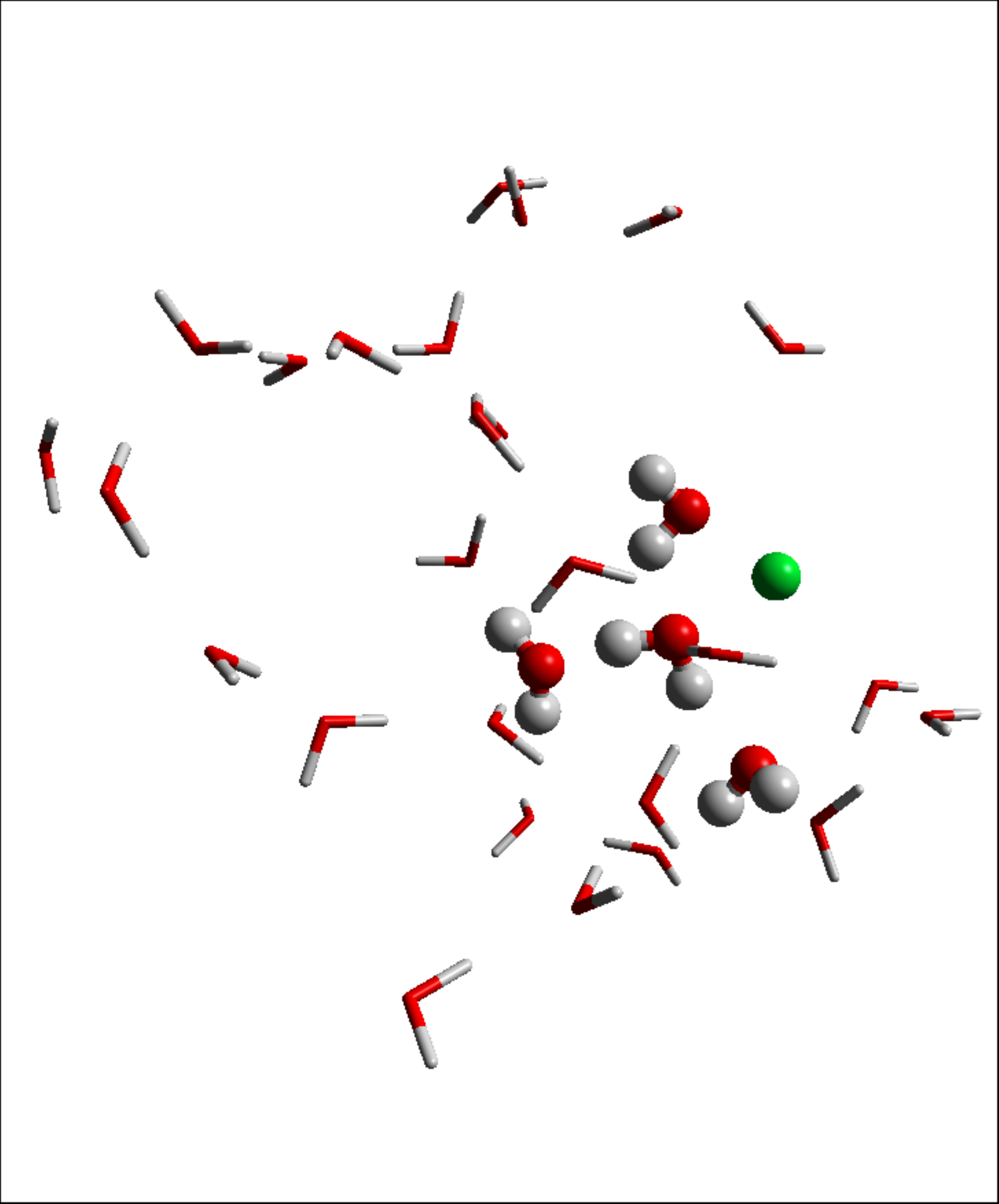}   \label{fig:dihyd1}}
\subfloat[88 fs] {\includegraphics[width=110pt]{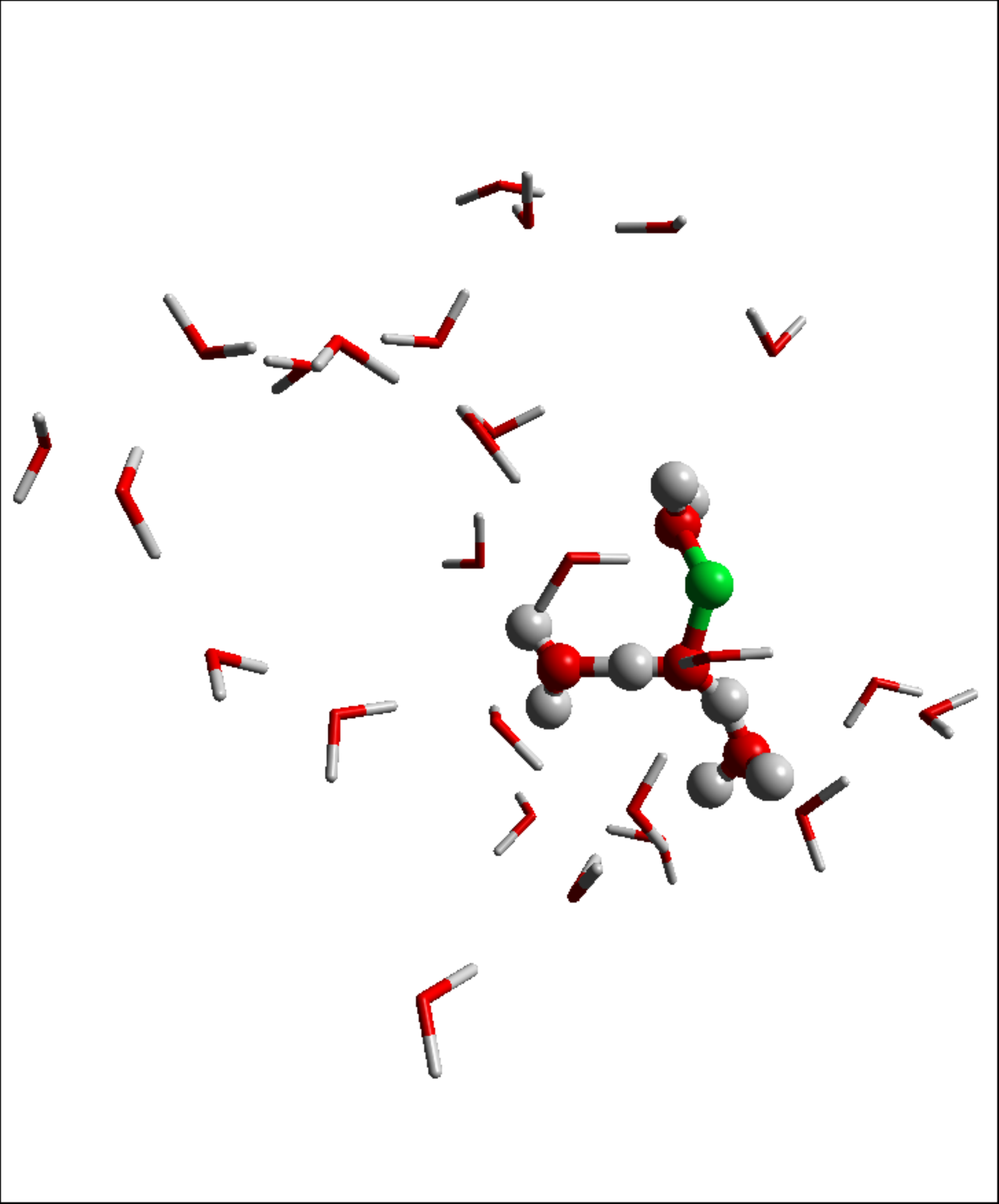}   \label{fig:dihyd2}}
\subfloat[109 fs]{\includegraphics[width=110pt]{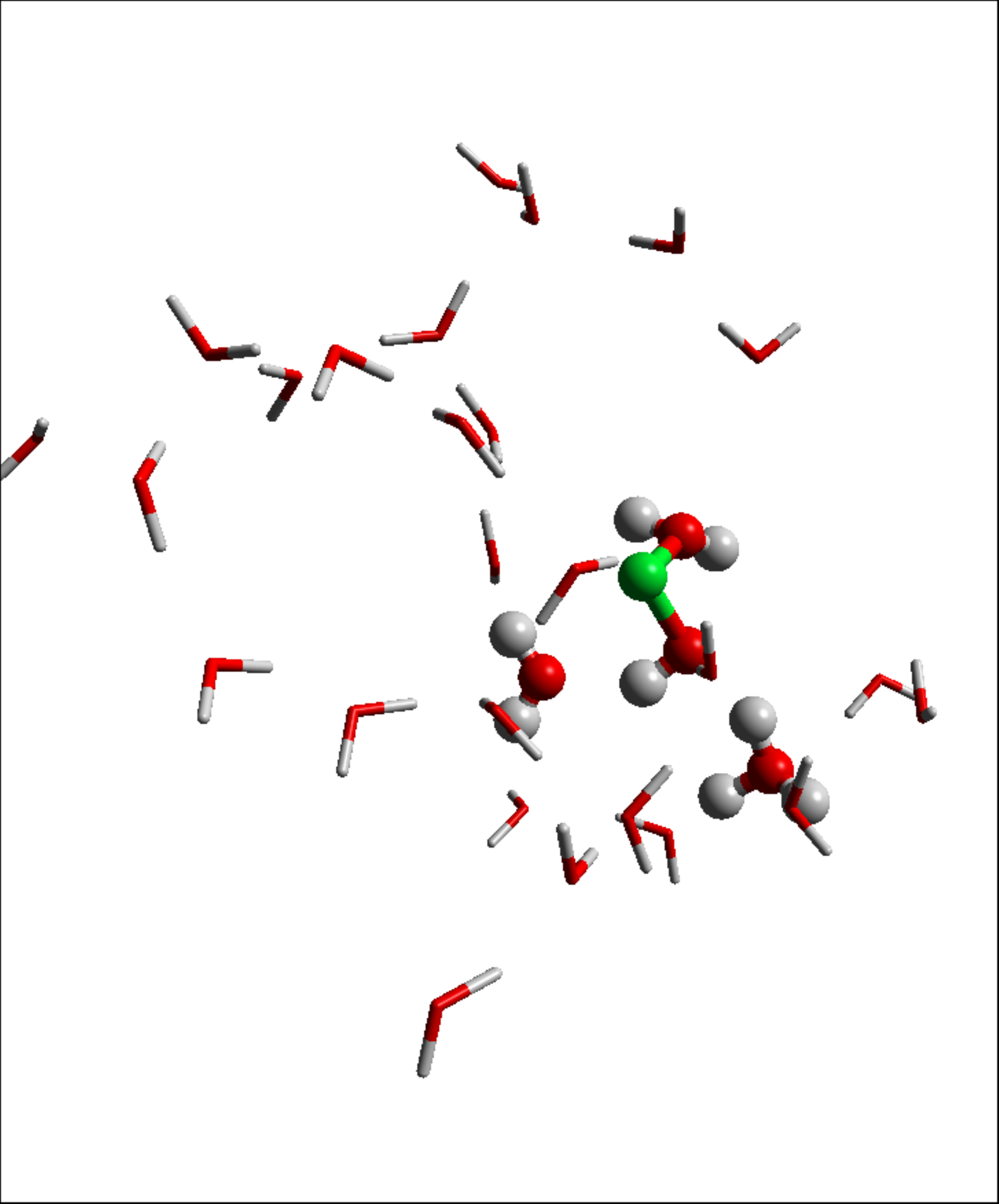}   \label{fig:dihyd3}}\\
\subfloat[210 fs]{\includegraphics[width=110pt]{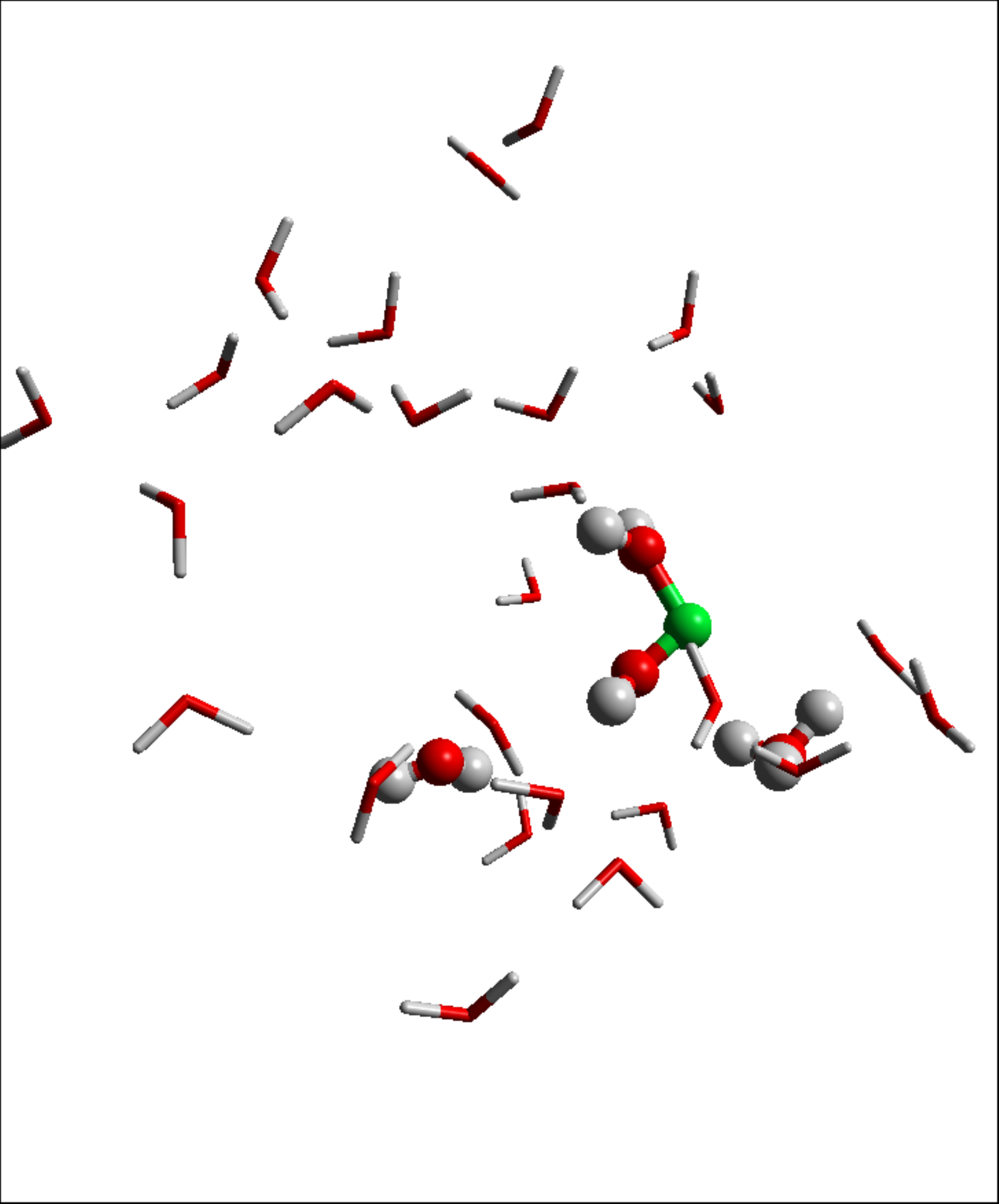}   \label{fig:dihyd4}}
\subfloat[271 fs]{\includegraphics[width=110pt]{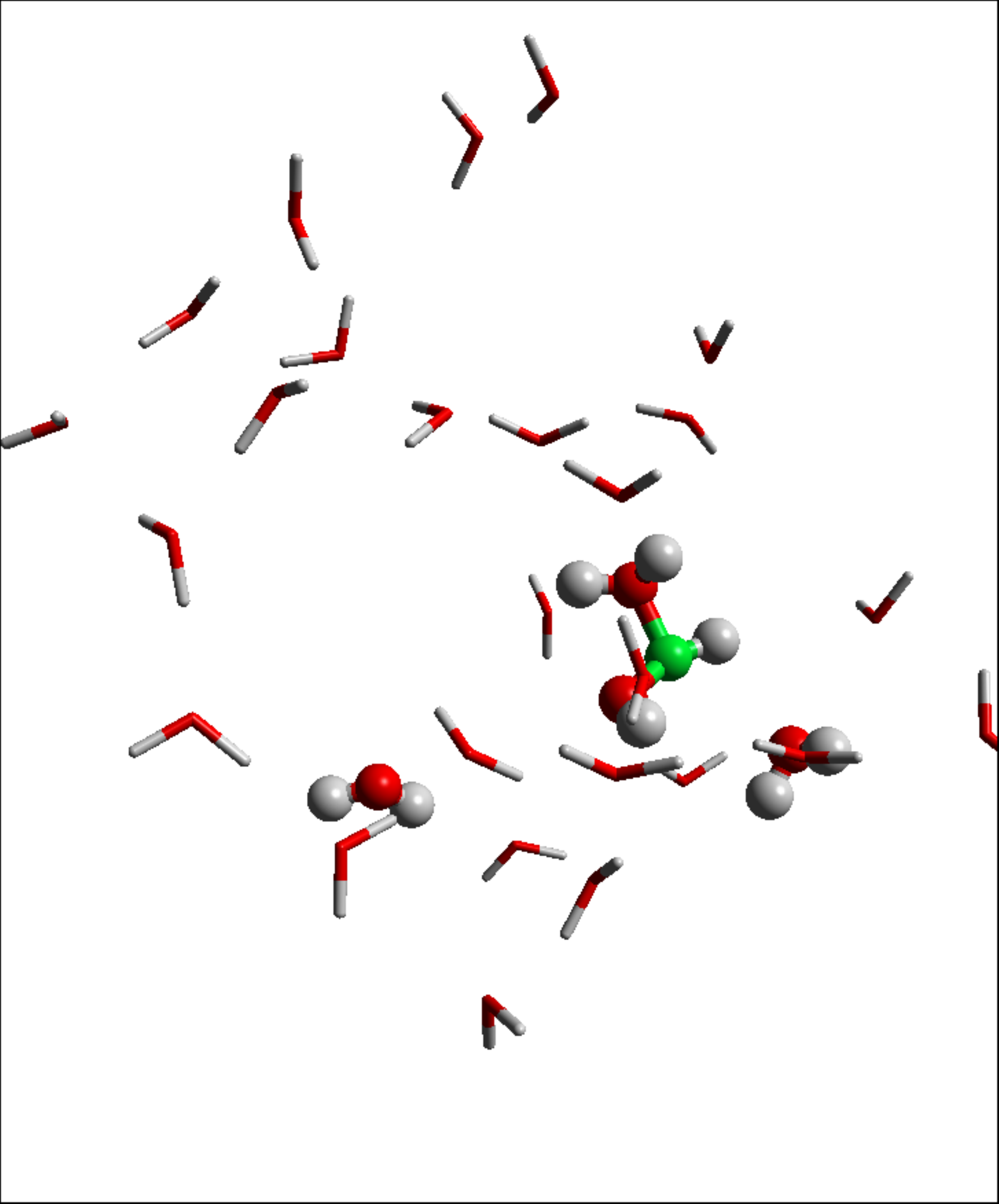}   \label{fig:dihyd5}}
\subfloat[447 fs]{\includegraphics[width=110pt]{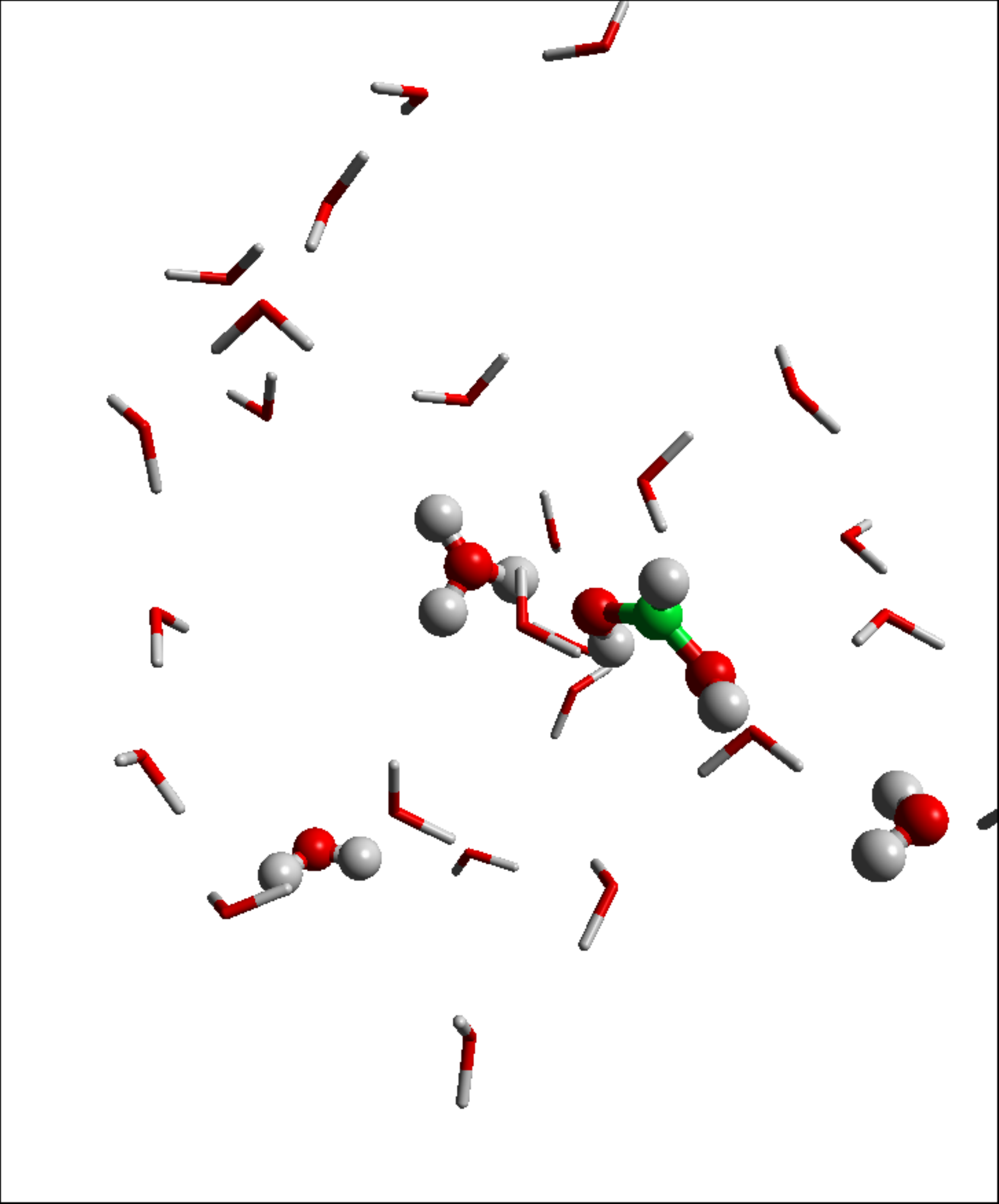}   \label{fig:dihyd6}}
\caption{A selection of frames from a trajectory in which the dihydroxymethyl 
radical (CH(OH)$_2\protect\rad$) forms. The atoms/molecules which participate directly 
in the reaction have been represented by \emph{spheres}. The remaining water molecules 
are shown as \emph{tubes}.}
\label{fig:3}
\end{figure}

A rare occurrence in these simulations is the creation of the dihydroxymethyl radical 
CH(OH)$_2\rad$, which is a relatively complex species involving a C atom and two water molecules.
Of the 30 trajectories simulated, three produced the CH(OH)$_2\rad$ species. In all cases the 
carbon ion had an initial energy of 1.7 eV with CH(OH)$_2\rad$ forming in the same manner each time, 
as depicted in \ref{fig:3}.

In this trajectory the carbon ion binds to a pair of water molecules through the 
oxygens. These, in turn, form hydrogen bonds with neighboring waters \emph{(b)}. 
Subsequently, a proton is removed from one of the waters in a similar manner to 
that seen in our previous trajectories. What remains is a carbon-water complex 
which now has a hydroxyl group (OH$^-$) bound to the carbon and a nearby hydronium 
\emph{(c)}. This carbon-bearing species now repositions itself over a time period 
of approximately 150 fs, making the carbon available to the hydronium \emph{(d)}. 
The carbon then accepts a proton from the hydronium ion \emph{(e)}. In the 
remainder of the simulation we observe that the carbon-bearing molecule loses 
a proton once more, leaving the neutral dihydroxymethyl radical and a hydronium ion
as final products.

Dihydroxymethyl is a precursor of more stable organic molecules such as formic acid (HCOOH), 
which is achieved by the loss of a hydrogen atom. The addition of a hydrogen atom, which
is an abundant species in the ISM, may produce methanediol (CH$_2$(OH)$_2$) which 
participates in the synthesis of simple sugars. 
Both formic acid and methanediol are predicted to form on grain surfaces and are of 
prebiotic significance \cite{Garrod2008}. Dihydroxymethyl was not proposed before because 
gas-phase work involved a single water molecule. These condensed-phase simulations 
highlight the possibility of near-neighbor interactions which are not a feature of 
gas-phase chemistry and provide a less complex path to chemical complexity. Some of these 
reactions have been observed in previous work with higher-energy C$^+$ projectiles 
interacting with water slabs \cite{Kohanoff2008} and are consistent with gas-phase work.

\section{Concluding remarks}

We have investigated the interaction of low-energy singly-charged carbon ions 
(C$^+$) with amorphous solid water clusters at 30 K via first-principles molecular 
dynamics simulations. We have selected initial projectile energies of 11 eV and 1.7 eV. 
These energies are attained after the (generally) more energetic projectiles 
(E $\sim$ 2 keV) \cite{Dawes2007} have travelled through the medium losing energy via 
inelastic collisions, as shown in a previous publication \cite{Kohanoff2008}. At such low 
energies the carbon ions can participate in chemical reactions and generate new organic 
chemical species. At the higher energy of 11 eV the kinetic energy of the carbon is large 
enough to dissociate a water molecule (5.1 eV) by impact alone, but not at the lower energy.
Along their way, the C$^+$ ions can also capture electrons produced by ionization thus
becoming neutral carbon atoms. In that case, the results of our previous study 
\cite{McBride2013} apply.

The dominant products are the isoformyl radical (COH$\rad$) and carbon monoxide (CO). 
COH$\rad$ formation is favored slightly among trajectories where the C$^+$ projectile 
has an initial energy of 1.7 eV. CO is produced more abundantly with 11 eV projectiles.
Depending on the energy of impact of the C$^+$ projectile, COH$\rad$ will be created 
via either a direct, ``knock-out'' mechanism, or through a COH$_2^+$ intermediate. In 
the direct mechanism the carbon ion will simply eject a proton from a water in a 
relatively rapid reaction.  This is more likely to occur at higher energies. The formation 
of the COH$_2^+$ intermediate occurs on a longer timescale. In each case the departing 
proton will be transferred among the water molecules of the cluster via a Grotthuss 
mechanism \cite{Grotthuss1806, Agmon1995} to form a hydronium ion. Carbon monoxide, 
the other major product of these 
reactions, is most commonly produced by the subsequent dissociation of COH$\rad$ into 
$\textrm{CO} + \textrm{H}\rad$, with the hydrogen atom becoming a secondary projectile 
that can react with other species in the sample, or simply evaporate.

A minor product of this reaction is the formyl radical (HCO$\rad$). This is formed most 
commonly following an impact of the carbon ion with a water which strips it of its 
hydrogens causing the creation of two hydronium species and CO$^-$. A proton then returns 
to the remaining carbon monoxide anion and attaches at the site of the carbon atom. 
HCO$\rad$ has also been produced via a hydroxymethylene 
intermediate. We observe more frequent production of the HCO$\rad$ radical here than in 
similar gas phase reactions. The availability of additional water molecules in the cluster 
allows for alternative routes to HCO$\rad$ formation.

If we compare this work with our previous study \cite{McBride2013} we see that formyl and 
isoformyl radicals along with carbon monoxide appear as products in both, although in the 
neutral case they are accompanied by H$\rad$ rather than the hydronium ions observed 
here. What we do not observe, however, is the hydroxymethylene radical which features prominently 
in our previous work and in the work of others who have studied the interaction of C with 
water \cite{Ahmed1983, Schreiner2006, Ozkan2012}.

An intriguing aspect is that we do not observe the formation of CO$_2$ as in experiments 
involving amorphous solid water, although the dihydroxymethyl radical may be a precursor 
of CO$_2$ on a longer timescale. Another possibility is that the COH$\rad$ reacts with an 
oxygen atom that has been produced by a higher energy projectile fully dissociating a 
water molecule, as observed in earlier work \cite{Kohanoff2008}. Yet another possibility is that the 
C$^+$ captures one electron in an excited state when approaching the sample at higher energy. 
This leaves H$_2$O$^+$, which dissociates as $\textrm{H}_2\textrm{O}^+ \rightarrow 
\textrm{OH}\rad + \textrm{H}^+$. The OH$\rad$ radical may then react with CO produced from 
a neutral carbon and water collision, as described in our previous publication \cite{McBride2013}, 
leading to $\textrm{CO} + \textrm{OH}\rad \rightarrow \textrm{CO}_2 + \textrm{H}\rad$. 
These are interesting avenues to explore in the future.

\acknowledgement
Astrophysics at QUB is supported by a grant from the STFC. EM was supported by DEL 
(Northern Ireland). The CP2K calculations were carried out at the HECToR facility
supported by EPSRC grants EP/F037325/1 and EP/K013459/1, allocated to the UKCP 
consortium.

\begin{suppinfo}
Results obtained using a dispersion-corrected functional.
\end{suppinfo}


\bibliography{carbon-ion-jpca}

\clearpage
\begin{figure}[ht]                                                                                   
\centering                                                                                            
\includegraphics[width=3.6cm]{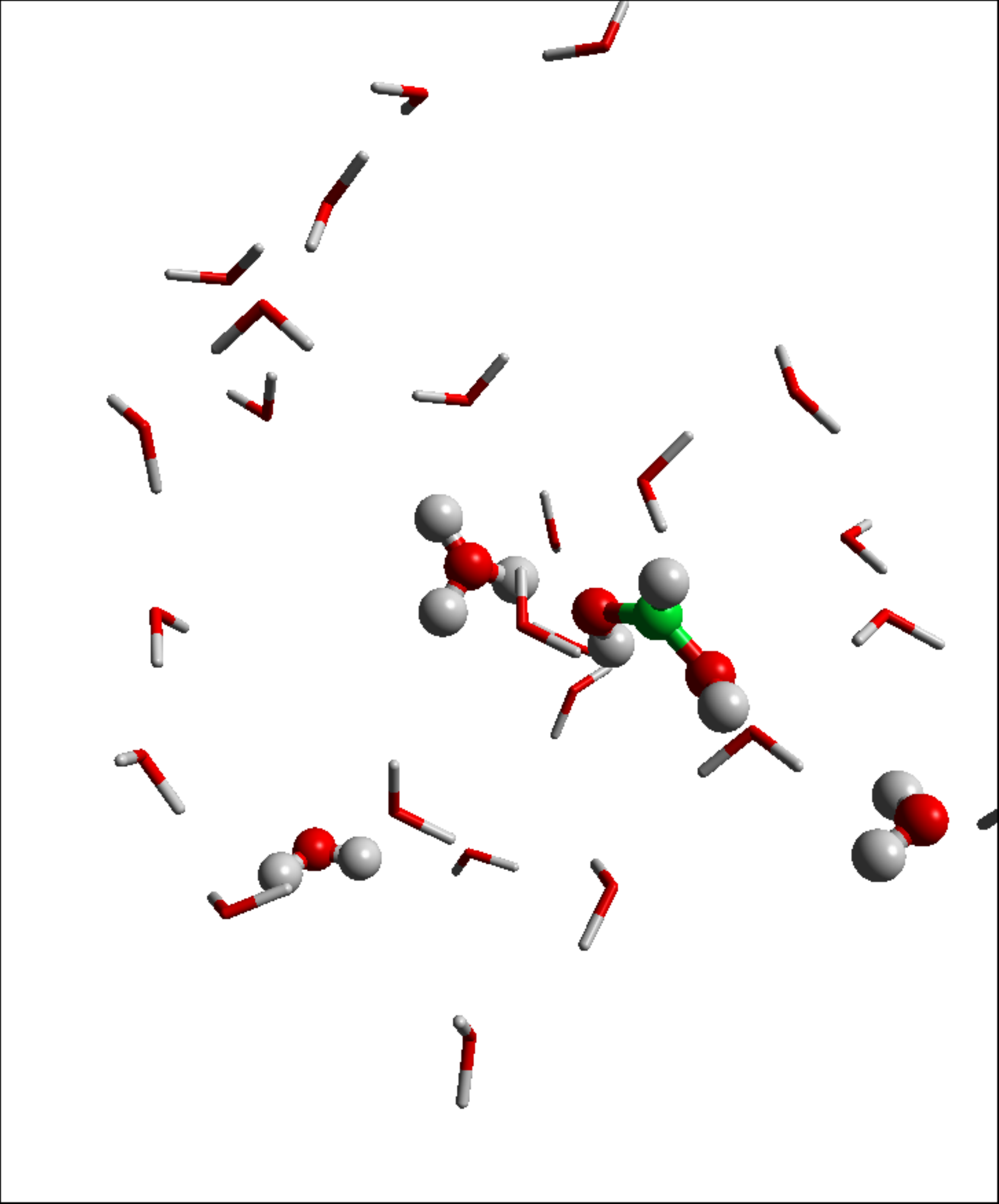}   
\label{fig:contents}                                                                                        
\end{figure}

\end{document}


\section*{Effect of dispersion interactions}

To assess the impact of the omission of dispersion forces from our work, a number of 
calculations were repeated using the PBE0 functional supplemented with Grimme's 
semi-empirical van der Waals interaction [S. Grimme, J. Comput. Chem. {\bf 27}, 1787 (2006)]. 

We found that when using the corrected functional the C-O bond length of the C$^+$-water 
complex does not change, remaining at 1.39 \AA. However, a small increase in the binding 
energy is produced, rising from 4.38 eV to 4.60 eV. To check dynamical properties, we
equilibrated our 30-molecule water cluster once again at 30 K, and run a few collision 
trajectories. Observation of trajectories and analysis of oxygen-oxygen distances 
indicated that the clusters share remarkably similar structures. The vast majority 
($\sim$ 94 \%) of the corrected distances are within 5 \% of their uncorrected counterparts. 

As a further check, a trajectory in which COH$\rad$ forms was repeated with the 
dispersion-corrected functional and it was found that COH$\rad$ was again produced. 
We also note that, in each trajectory, key events in the molecule's formation mechanism 
occur practically at the same time (within a few fs) when comparing corrected and 
uncorrected functionals.